\documentclass[review,onecolumn,12pt]{elsarticle}
\usepackage[margin=1in]{geometry} 

\usepackage{pifont}

\usepackage{soul}  
\usepackage{xcolor}
\usepackage[dvipsnames]{xcolor}
\usepackage[table]{xcolor}
\usepackage{adjustbox}
\usepackage{subcaption}

\usepackage{amssymb}

\usepackage[flushleft]{threeparttable}
\usepackage{multirow}
\usepackage[utf8]{inputenc}
\usepackage{enumerate} 
\usepackage{graphicx}
\usepackage{textgreek}
\usepackage{textcomp}
\usepackage{listings}
\usepackage{array}
\usepackage[switch,mathlines]{lineno}
\usepackage{xfrac}
\usepackage{graphicx}
\biboptions{square,numbers,sort&compress}
\usepackage{mathtools}
\usepackage{wrapfig}
\usepackage{enumitem}
\usepackage{cuted}
\usepackage{algorithmicx}%
\usepackage{algorithm,algpseudocode}
\algnewcommand{\Inputs}[1]{%
  \State \textbf{Inputs:}
  \Statex \hspace*{\algorithmicindent}\parbox[t]{.8\linewidth}{\raggedright #1}
}
\algnewcommand{\Initialize}[1]{%
  \State \textbf{Initialize:}
  \Statex \hspace*{\algorithmicindent}\parbox[t]{.8\linewidth}{\raggedright #1}
}

\usepackage{algpseudocode}

\usepackage{stackengine}

\usepackage{CJK}
\usepackage{bm}
\usepackage{color}
\usepackage{amsmath,amsthm,amssymb,amsfonts}
\usepackage{flushend}
\usepackage{float}
\usepackage{multirow}

\newtheorem*{remark*}{Remark}
\usepackage{arydshln} 
\usepackage{hyperref}
\hypersetup{
     colorlinks   = true,
     linkcolor    = blue,
     citecolor    = red
}

\usepackage[colorinlistoftodos]{todonotes}

\makeatletter
\newcommand*{\transpose}{%
  {\mathpalette\@transpose{}}%
}
\newcommand*{\@transpose}[2]{%
  \raisebox{\depth}{$\m@th#1\intercal$}%
}
\makeatother

\makeatletter
\g@addto@macro\normalsize{%
\setlength\abovedisplayskip{3pt}
\setlength\belowdisplayskip{3pt}
\setlength\abovedisplayshortskip{3pt}
\setlength\belowdisplayshortskip{3pt}
}
\makeatother

\usepackage{soul}
\usepackage{tikz}
\usepackage{booktabs}
\usepackage{tabularx}

\journal{Ad Hoc Networks}

\let\oldbibliography\thebibliography
\renewcommand{\thebibliography}[1]{%
  \oldbibliography{#1}%
  \setlength{\itemsep}{1.9pt}%
  \setlength{\baselineskip}{12pt}
  \setlength{\lineskiplimit}{-\maxdimen}
}

\begin{document}

\begin{frontmatter}

\title{Federated Lightweight Intrusion Detection in Drone Swarms with Knowledge Distillation}
\author[]{Fawaz J.~Alruwaili}
\author[]{Cihan Tunc}
\address{Department of Computer Science and Engineering, University of North Texas, Denton, TX 76201 USA}
\ead{FawazAlruwaili@my.unt.edu, cihan.tunc@unt.edu} 

\begin{abstract}
Drone swarms are increasingly deployed in critical applications such as surveillance, disaster response, and infrastructure monitoring. However, their reliance on open communication channels and their limited computational resources make them vulnerable to a wide range of cyber-threats. There is a growing interest in intrusion detection systems (IDS) specifically designed for drone environments and operations. However, the conventional solutions including Machine Learning (ML)-based approaches require collecting all data from heterogeneous drones in the swarm and processing on a central server may not be always feasible. 
Federated Learning (FL) has emerged as a promising distributed solution with an additional privacy-preserving feature. 
Even though potential studies exist, conventional FL-based IDS frameworks still face communication and computational overhead challenges, while achieving a balance between efficiency and effective detection under practical resource constraints remains a challenge. 
Therefore, we propose a lightweight FL-based IDS tailored for drone swarm networks using deep neural networks (DNN) enhanced with knowledge distillation (KD) to reduce model complexity and communication costs without sacrificing detection performance. We evaluate our framework using Raspberry Pi 4 devices and a real-world drone network dataset. 
Our approach demonstrates a detection accuracy of approximately 98.6\% while reducing overall communication cost by around 70\% and computational overhead by 29\%. These results show that FL combined with KD is a practical and suitable solution for secure and efficient deployment in resource-constrained drone networks.
\end{abstract}

\begin{highlights}

\item Contribution 1. Federated Learning based intrusion detection system for drone swarms. 
\item Contribution 2. Knowledge distillation to reduce model complexity, communication, and computation costs. 
\item Contribution 3. Physical demonstration of the proposed approach with Raspberry Pi 4 and a real-world drone network dataset. 

\end{highlights}


\end{frontmatter}

\newcommand{\xmark}{\ding{55}}

\section{Introduction}
Drones are becoming a key component across various sectors such as surveillance, transportation, disaster management, agriculture, and infrastructure monitoring due to their mobility, compact size, low cost, ease of deployment, and reduced infrastructure requirements~\cite{ghosh2024flas, lan2025application, tauseef2025comprehensive}. 
For example, in the agricultural sector, drones are employed for crop monitoring, irrigation management, disease detection, and yield forecasting~\cite{elhesasy2024drone}, with the agricultural drone market expected to grow from approximately \$1 billion in 2022 to over \$5 billion by 2030~\cite{singh2024drone}. 
Furthermore, as of June 2026, Federal Aviation Administration (FAA) reported 837,512 registered drones in the United States alone~\cite{FAA2026numbers}.
Despite advantages, individual drone operations are restricted due to the limited onboard resources including battery capacity, computational resources, communication bandwidth, as well as possible failures and physical attacks/obstacles~\cite{krichen2025lightweight, mukherjee2023overview, thai2024recent}.
Hence, drone swarms have been utilized to enhance a single drone's ability to perform missions in a collaborative manner. 

Even though drone and drone swarm operations are gaining popularity, the complex coordination and communication requirements introduce vulnerabilities, such as man-in-the-middle (MITM) attacks, distributed denial-of-service (DDoS) attacks, and GPS spoofing~\cite{yang2024fed}. 
For example, MITM attack can compromise the availability, integrity, or confidentiality of data exchanged between drones or between a drone and the ground control station (GCS), where an attacker can either intercept the communication channel to capture sensitive data without altering or intercepting by modifying the data exchanged~\cite{branco2025cyber}. 
In addition to MITM attacks, DoS attacks represent another critical threat, aiming to make a system/service temporarily inaccessible by overwhelming its resources, hence preventing the target drone from functioning and disrupting overall network availability~\cite{branco2025cyber, maham2024new}. 

Considering the future with a large number of drones deployed as a part of swarms, there is a need for robust drone swarm intrusion detection systems (IDS) for reliable and efficient drone swarm operations. Standalone IDS solutions are not suitable/applicable as they are not built for drones~\cite{tufekci2024dude} and may have outdated or insufficient data~\cite{hadi2024real}, which restricts their ability to detect threats effectively or cannot be run on drones with limited power and computational resources. This can cause a single compromised drone to expose an entire drone swarm network to significant security risks. 
Additionally, considering number drones in swarms and volume of data generated and transferred, the traditional centralized IDS solutions become questionable~\cite{fan2023uav, deng2025fidsus} due to high communication overhead, latency, and potential network congestion~\cite{poorvi2025reliable, cal2024energy} as well as network connectivity~\cite{mukherjee2023overview}. 
These drawbacks highlight the need for cooperative IDS approaches that enable data sharing and collective threat intelligence to improve detection accuracy and accelerate response to emerging threats. 

We propose addressing the drone swarm IDS problem using federated learning (FL) for collaborative drone learning by enabling decentralized model training across distributed drone clients and sharing model weights to a centralized server rather than complete raw data. This enables us to reduce communication overhead, mitigate privacy concerns, and improve performance with efficiency compared to centralized approaches~\cite{din2025federated, poorvi2025reliable, yang2024fed, li2023efficient}. 
Nevertheless, drone swarms can consist of heterogeneous drones with different configurations and capabilities, bringing additional challenge.  
Recent studies suggest contribution quantification and affinity-based feature aggregation to address client heterogeneity and communication constraints~\cite{yang2024fed, deng2025fidsus}, asynchronous FL models to mitigate latency and communication overhead by leveraging drone-based model aggregation~\cite{zhai2024uav}, etc. 

This work addresses the trade-offs between security, efficiency, and resource constraints in drone swarm networks by an FL-based IDS incorporating knowledge distillation (KD). By enabling participating drones to train lightweight models and by minimizing transmitted model sizes, our approach significantly reduces both communication and computational overheads  while maintaining accuracy and overall model performance. 
The contributions of this paper are as follows:
\begin{enumerate}
    \item We propose a FL-based IDS framework designed for heterogeneous drone swarm networks, addressing the unique challenges of limited computational resources, communication overhead, and security in drone environments. 
    
    \item We introduce a server-side KD strategy that refines aggregated student models using a pre-trained teacher model and a proxy dataset, reducing both communication and computational overhead while improving model performance.
    
    \item We implement and evaluate our proposed framework on real-world resource-constrained hardware (Raspberry Pi 4 devices) using a real-world drone network dataset, demonstrating its effectiveness in balancing accuracy, efficiency, and adaptability in heterogeneous drone swarm scenarios.
\end{enumerate}

To clarify the assumptions and attack surface considered in this work, we define our threat model in Section~\ref{sec:related_work}, along with related work in this domain. In Section~\ref{sec:methodology}, we present the proposed methodology and the steps needed to achieve FL-based IDS for drone swarm networks. Section~\ref{sec:results} details the conducted experiments and their results. Finally, the paper is concluded in Section~\ref{sec:conclusion}.

\section{Background and Related Works} 
\label{sec:related_work}
The security of drone swarms is increasingly viewed as a multi-layered challenge, requiring defenses at both the physical and network levels. 
Some of the prior work has explored a range of control-theoretic and sensor-level attack detection mechanisms for drone systems. For instance, invariant-based methods detect anomalies by monitoring violations of physical system constraints~\cite{ zhong2025robin}, while MMIO-based approaches, such as M2MON~\cite{Arslan2021} identify attacks by analyzing low-level memory-mapped I/O access patterns. In addition, Kalman filter-based techniques~\cite{elsayed2025combination} have been used to detect sensor manipulation and signal injection attacks by modeling system state estimation errors. These approaches primarily focus on detecting physical-layer and sensor-level attacks affecting control dynamics and onboard measurements. 

In modern deployments, drones communicate not only through onboard control protocols but also via IP-based wireless networks (e.g., WiFi, LTE, or mesh networks) or radio signals to exchange telemetry, coordination messages, and data with other drones and GCS~\cite{chavekar2023review, chen2025development, maeng2023lte, mushtaq2024framework}. 
An adversary is assumed to be capable of observing, injecting, or manipulating network traffic within these communication channels. Specifically, the threat model considers representative network-layer attacks, including DoS/DDoS attacks (e.g., TCP SYN and UDP flooding) and MITM attacks, which aim to disrupt communication, exhaust network resources, or compromise data integrity~\cite{branco2025cyber, adhikari2025unmanned, chavekar2023review, tufekci2024dude}. 
While low-level communication protocols such as MAVLink are commonly used between onboard components and flight controllers~\cite{wray2025pave}, many real-world drone systems rely on higher-layer network communication for coordination and data exchange, particularly when integrated with edge computing platforms or GCS~\cite{adhikari2025unmanned, chen2025development, chavekar2023review, mushtaq2024framework}. Therefore, this work targets the network communication layer, which remains vulnerable to traffic-based attacks~\cite{adhikari2025unmanned}.

Focusing on the communication layer, several studies have explored FL-based drone network IDS using different optimization techniques. For example, a series of studies~\cite{cal2024energy,yao2025resource} proposed the FedKD framework, an energy-efficient federated KD framework designed for the Internet of Drones (IoD), where each drone trains both a teacher and a student model, but only the student model is sent to the server, which reduces communication overhead. The initial study~\cite{cal2024energy} focused on optimizing CPU frequencies to minimize energy consumption by formulating a non-linear programming problem that adapts each drone's processing speed based on its distance to the server, data volume, and hardware limits. The framework was extended in a later study~\cite{yao2025resource} by integrating the optimization of CPU resources, wireless transmission power, and bandwidth allocation through an iterative algorithm. Evaluations on the MNIST dataset~\cite{deng2012mnist} using a Convolutional Neural Network (CNN) showed that the enhanced method (FedKD-FPB) achieved approximately 98\% accuracy and reduced energy consumption by up to 85\% compared to FedKD and 94\% compared to FedAVG.

Deng et al.~\cite{deng2025fidsus} proposed FIDSUS, a federated IDS framework for drone swarms employing maximum mean discrepancy (MMD) to detect changes in a client's data distribution over time and down-weight unstable updates. It also employs affinity matrices to select and aggregate feature extractors from the most similar clients, which reduces communication overhead by transmitting only classifier parameters and class-wise feature representations rather than full model updates. The framework was evaluated on NSL-KDD and UNSW-NB15 datasets with three different numbers of drone clients (10, 50, 100) and compared with seven FL methods: FedAvg, FedProx, MOON, FedAvgDBE, FedProto, GPFL, and FedGH employing a 1D CNN model. The proposed FIDSUS achieves the highest accuracy across both datasets, reaching up to 97\% accuracy on NSL-KDD with NC=10 and 89\% on UNSW-NB15 with 10 clients. Alternatively, the framework introduced in~\cite{ihekoronye2023federated} proposed utilizing local differential privacy (LDP) as a privacy-preserving technique by adding noise to model updates to protect sensitive information against inference attacks. A DNN model was trained on the Edge-IIoT dataset~\cite{ferrag2022edge}, achieving up to 90\% accuracy using Laplace noise with $\varepsilon = 0.3$, and maintained high performance even when scaling from 15 to 50 clients.

Beyond drone security domain, federated IDS has also been explored in IoT and edge environments. For example, a study~\cite{li2023efficient} proposed DAFL, an efficient FL-based IDS designed to improve both accuracy and communication efficiency. In this framework, clients with poor local detection performance are excluded from aggregation, and dynamic weighted averaging is applied based on both sample size and model quality. These strategies enable DAFL to converge faster and reduce the number of communication rounds. The proposed framework was evaluated using a CNN model on the CSE-CIC-IDS2018 dataset~\cite{sharafaldin2018toward}, achieving an average of 94.6\% across accuracy, precision, recall, and F1-score, while reducing communication overhead by 33\% to 71\% compared to baseline FL. Another study~\cite{yang2024fed} proposed Fed-FIDS, an efficient FL-based IDS framework design for edge environments with heterogeneous clients with three key components: a real-time contribution quantification module that evaluates how much each client's update improves the global model's performance, selecting clients with the highest contribution for future training rounds, resource allocation module for allocating more resources (e.g., longer training time or better bandwidth) to clients who historically contribute more, and a Data Selection Module to determine how much local data should use in each round considering the client's computing ability and assigned resources. The proposed Fed-FIDS was evaluated on the UNSW-NB15 and Edge-IIoT datasets using DNN, 1D-CNN, and 1D-CNN-LSTM models, achieving accuracy improvement 0.83\% to 4.98\% on the UNSW-NB15 and 0.10\% to 0.47\% on the Edge-IIoT dataset and a $144\times$ reduction in local training time compared to baseline approaches. 

A summary of these related studies, including models, datasets, performance, efficiency considerations, and validation environments, is presented in Table~\ref{tab:comparison_summary}. 
Although existing studies provide valuable advancements in FL-based IDS, they do not address the challenge and necessity of balancing security, efficiency, and model performance in drone networks. These trade-offs are often amplified by the resource-constrained nature of drones and the communication demands of distributed learning.

\begin{table*}[ht]
\centering
\caption{Summary of related work illustrating efficiency considerations, dataset relevance, and deployment strategy.}
\renewcommand{\arraystretch}{1.2}
\adjustbox{max width=\linewidth}{

\begin{tabular}{|c|c|c|c|c|c|c|c|c|c|}

\hline
\textbf{Study} & \textbf{Year} & \textbf{Model} & \textbf{Dataset} & \begin{tabular}[c]{@{}c@{}}\textbf{Drone}\\\textbf{Relevancy}\end{tabular} & \begin{tabular}[c]{@{}c@{}}\textbf{Dataset}\\\textbf{Year}\end{tabular} & \textbf{Accuracy} & \begin{tabular}[c]{@{}c@{}}\textbf{Resource}\\\textbf{Efficiency}\end{tabular} & \begin{tabular}[c]{@{}c@{}}\textbf{Comm}\\\textbf{Efficiency}\end{tabular} & \textbf{Validation} \\
\hline \hline 
\cite{cal2024energy} & 2024 & \multirow{2}{*}{CNNs} & \multirow{2}{*}{MNIST} & \multirow{2}{*}{\xmark} & \multirow{2}{*}{1998} & \multirow{2}{*}{98\%} & \multirow{2}{*}{\checkmark} & \multirow{2}{*}{\checkmark} & \multirow{2}{*}{Simulation} \\
\cline{1-2}
\cite{yao2025resource} & 2025 & & & & & & & & \\
\hline
\cite{deng2025fidsus} & 2025 & 1D CNN & \begin{tabular}[c]{@{}c@{}}UNSW-NB15\\NSL-KDD\end{tabular} & \xmark & \begin{tabular}[c]{@{}c@{}}2015\\2009\end{tabular} & \begin{tabular}[c]{@{}c@{}}89\%\\97\%\end{tabular} & \xmark & Partially & Simulation \\
\hline
\cite{ihekoronye2023federated} & 2023 & DNN & edge-IIoT & \xmark & 2022 & 90\% & \xmark & \xmark & Simulation \\
\hline
\cite{li2023efficient} & 2023 & CNN & CSE-CIC-IDS2018 & \xmark & 2018 & 94\% & \xmark & \checkmark & Simulation \\
\hline
\cite{yang2024fed} & 2024 & \begin{tabular}[c]{@{}c@{}}DNN\\1D-CNN\end{tabular} & \begin{tabular}[c]{@{}c@{}}UNSW-NB15\\Edge-IIoT\end{tabular} & \xmark & \begin{tabular}[c]{@{}c@{}}2015\\2022\end{tabular} & \begin{tabular}[c]{@{}c@{}}79\%\\95\%\end{tabular} & \checkmark & \checkmark & Simulation \\
\hline
\textbf{Ours} & \textbf{2025} & \textbf{DNN} & \begin{tabular}[c]{@{}c@{}}\textbf{ISOT Drone}\\\textbf{Dataset}\end{tabular} & \checkmark & \textbf{2024} & \textbf{98\%} & \checkmark & \checkmark & \begin{tabular}[c]{@{}c@{}}\textbf{Real hardware}\\\textbf{deployment}\end{tabular}  \\ 
\hline
\end{tabular}
}
\label{tab:comparison_summary}
\end{table*}

\textit{Despite meaningful progress, several common limitations persist across existing FL-IDS frameworks for drone environments.} For example, the frameworks in~\cite{cal2024energy,yao2025resource} require each drone to host both teacher and student models, which increases local computation and memory usage. This conflicts with the resource-constrained nature of drones. Similarly, the FIDSUS framework~\cite{deng2025fidsus} adds client-side overhead by requiring affinity scoring and the storage of historical feature representations for each class. While local differential privacy in~\cite{ihekoronye2023federated} enhances privacy by adding noise to each client's model updates, it introduces additional computation due to noise generation and modified model updates, which may further overload drone resources and can also reduce model accuracy and make training less stable. Adding noise can also hide important patterns, reducing the benefits of IDS and collaborative learning. Similarly, selecting clients with the highest contribution~\cite{yang2024fed} or excluding low-performing clients~\cite{li2023efficient} may bias the system toward those with clean or common data, while overlooking rare but valuable patterns. Likewise, reducing the influence of clients based on distribution deviation metrics, such as MMD in~\cite{deng2025fidsus}, may unintentionally ignore valid but unusual activity, creating blind spots or delaying the detection of novel intrusions. 

In addition to the aforementioned concerns, many existing FL-IDS frameworks are limited by their reliance on general-purpose datasets and lack of real-world hardware validation. For instance, studies such as~\cite{cal2024energy,yao2025resource, ihekoronye2023federated, yang2024fed, li2023efficient, deng2025fidsus} employ datasets developed for broader IoT or network intrusion scenarios, such as MNIST, Edge-IIoT, CSE-CIC-IDS2018, NSL-KDD, and UNSW-NB15. While these datasets are widely used, some are outdated or contain insufficient drone-relevant traffic patterns, making them less suitable for capturing the unique communication behavior and threat landscape of drone networks. Furthermore, evaluations in~\cite{cal2024energy,yao2025resource, ihekoronye2023federated, li2023efficient, deng2025fidsus} were conducted in simulated environments without deployment on actual hardware, leaving questions about real-world performance, system overhead, and deployability under practical drone conditions. \textit{In contrast,} our proposed framework addresses these limitations by employing KD and offloading computationally intensive tasks to the server, allowing drones to operate only lightweight student models. This design reduces computation and communication costs while preserving accuracy. Moreover, we evaluate our framework using the ISOT Drone dataset~\cite{chen2025drone}, which is a recent and large-scale dataset specifically designed for drone intrusion scenarios using real Raspberry Pi hardware, demonstrating a practical and effective solution for secure and efficient IDS in drone swarms.

\section{Methodology} \label{sec:methodology}

\subsection{System Architecture}

\begin{figure*}[bh]
\centering
\includegraphics[width=\linewidth]{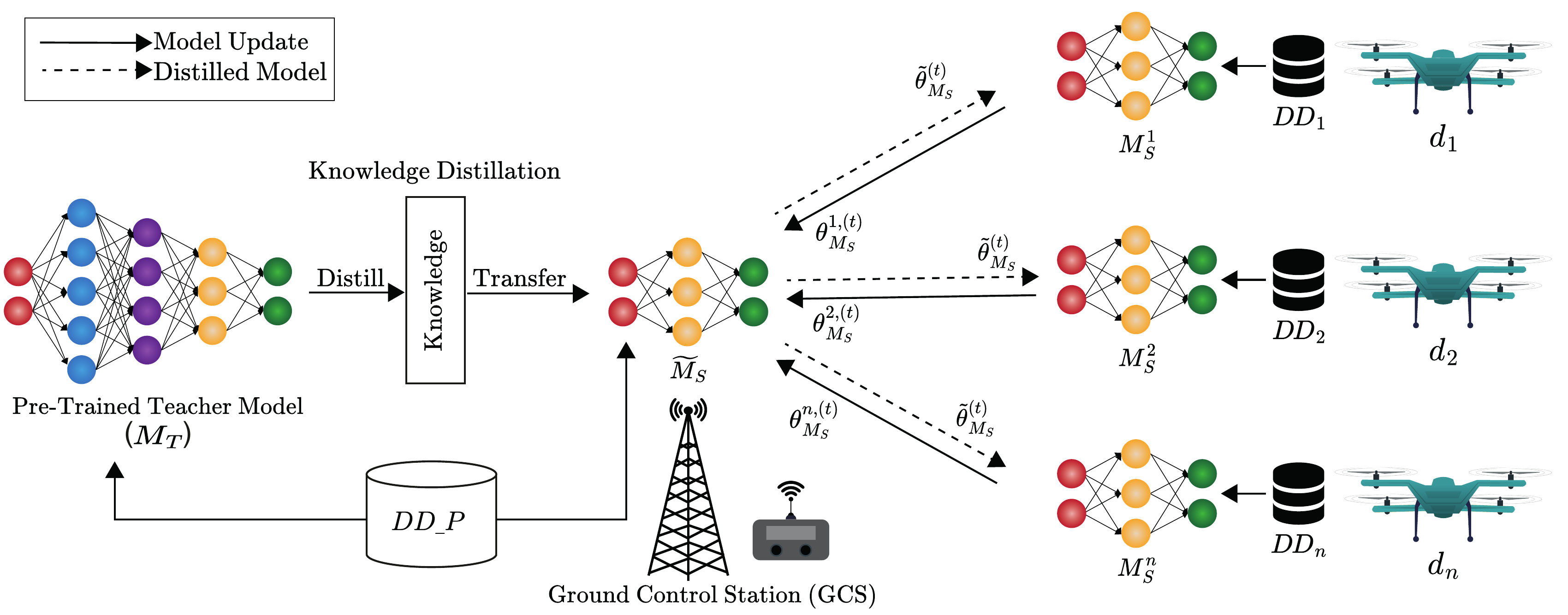}
\caption{Proposed FL-KD approach architecture.}
\label{workflow}
\end{figure*}

The proposed FL-based IDS using knowledge distillation (FL-KD) architecture  for drone swarm network is illustrated in~Fig.~\ref{workflow}. A group of drones (as a swarm) operates collaboratively, where each drone acts as an edge client and the ground control station (GCS) serves as the central server to enable secure and efficient collaborative learning while preserving data privacy, minimizing resource consumption, and maintaining overall model performance. 

Let $\mathcal{D}$ denote the drone swarm consisting of drones $d_i \in \mathcal{D}$. Each drone $d_i$ possesses its own local dataset $dd_i$ containing network traffic data collected onboard, which remains entirely private and never shared externally. The training process proceeds through FL rounds denoted by $r$, which indicates the current communication round. The local `student' model for drone $d_i$ is denoted by $M_S^i$, with parameters (weights) $\theta_{M_S}^{i,(r)}$ at round $r$. After local training on $dd_i$ for drone $d_i$ completed at round $r$, the updated weights $\theta_{M_S}^{i,(r+1)}$ are sent to the server (i.e., GCS) for the $(r+1)^{th}$ iteration (i.e., next iteration) corresponding to the updated model state to be aggregated by the GCS to form the global student/drone model $\widetilde{M}_S$ for the next round with the aggregated weights of $\widetilde{\theta}_{M_S}^{(r+1)}$. The global model is then distributed back to all drones for the next training iteration. 
To further enhance the generalization and performance of the global student model, the GCS performs server-side KD using a pre-trained teacher model $T_M$ on a proxy drone dataset $DD\_P$, which consists of previously selected subset network traffic data stored at the server. The proxy drone dataset $DD\_P$ is distinct from the local datasets $dd_i$ held privately by each drone, and it is used to generate soft labels for improving the student model's generalization. These soft labels help to refine the aggregated global student model through distillation loss.

\subsection{Operational Flow}
\begin{figure}[ht]
\centering
\includegraphics[width=.7\linewidth]{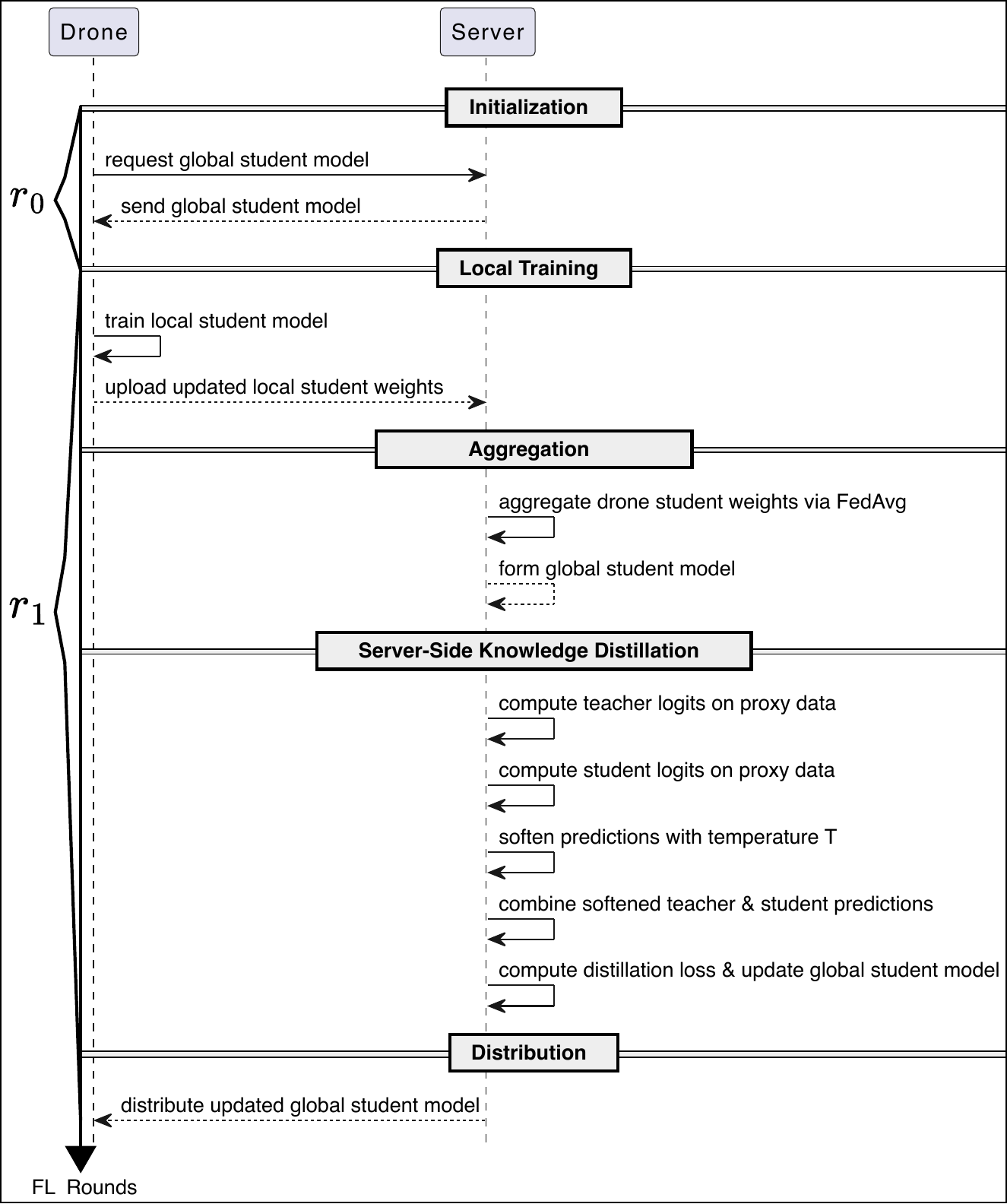}
\caption{Proposed FL-KD operational sequence diagram.}
\label{uml}
\end{figure}

The operational flow of our proposed FL-KD approach is depicted in Fig.~\ref{uml}, highlighting the interaction between the drones and the server and how local updates and server-side KD collectively enhance the global student model for the drone swarm IDS.

At the beginning, at the initial round~$r_0$, the server initializes the global student model, representing the untrained weights. The model is then distributed to all participating drones $d_i \in \mathcal{D}$. Each drone sets its local student model $M_S^i$ to this initial version, as defined in Eq.~\ref{eq:init_model}, 
\begin{equation}
\theta_{M_S}^{i,(0)} \leftarrow \theta_{M_S}^{(0)}
\label{eq:init_model}
\end{equation}
where~$\mathcal{D}$ denotes the set of all drones participating in the swarm, $\theta_{M_S}^{i,(0)}$ denotes the weights of the local student model at drone~$d_i$ during round~$r_0$, and~$\widetilde{\theta}_{M_S}^{(0)}$ is the initial weights of the global student model before any training. 
Then, each drone trains its local model $M_S^i$ on its own network traffic data $DD_i$ by minimizing the categorical cross-entropy loss between predictions and hard labels $Y_i$, as in Eq.~\ref{eq:local_training},
\begin{equation}
\theta_{M_S}^{i,(1)} = \arg\min_{\theta} \mathcal{L}_{\text{CE}}\left(Y_i, M_S(DD_i; \theta)\right)
\label{eq:local_training}
\end{equation}
where $DD_i$ is the local dataset at drone $d_i$, $Y_i$ are the corresponding labels, and $M_S$ is the local student model trained on drone $d_i$. The notation $\theta_{M_S}^{i,(1)}$ indicates the updated weights after local training during the first FL round $r_1$.

After local training, each drone sends its updated model $\theta_{M_S}^{i,(1)}$ to the server. The server aggregates the received student models to form the global student model using the standard Federated Averaging (FedAvg) algorithm, as defined in Eq.~\ref{eq:fedavg},
\begin{equation}
\theta_{M_S}^{(1)} = \frac{1}{|\mathcal{D}|} \sum_{i \in \mathcal{D}} \theta_{M_S}^{i,(1)}
\label{eq:fedavg}
\end{equation}
The aggregated global student model $\theta_{M_S}^{(1)}$ integrates knowledge from all drones without exposing raw data. To further refine it, the server performs a KD technique using a pre-trained teacher model $M_T$ and a public proxy dataset $DD\_P$. The logits from both models on $DD\_P$ are computed in Eq.~\ref{eq:logits}:
\begin{equation}
Z_T = M_T(DD\_P), \quad Z_S = M_S^{(1)}(DD\_P)
\label{eq:logits}
\end{equation}
Then, the KD loss function incorporates both cross-entropy with ground-truth labels and a softened KL divergence between the teacher and student logits, as in Eq.~\ref{eq:kd_loss},
\begin{multline}
\mathcal{L}_{\text{KD}} = \alpha \cdot \mathcal{L}_{\text{CE}}(Y_P, Z_S) + (1 - \alpha) \cdot T^2 \cdot \text{KL}\left( 
\text{softmax}\left(\frac{Z_T}{T}\right) \,\|\, 
\text{softmax}\left(\frac{Z_S}{T}\right) \right)
\label{eq:kd_loss}
\end{multline}
where $T$ is the temperature parameter that controls the softness of the probability distributions across classes, and $\alpha \in [0,1]$ balances the hard and soft labels losses.

The categorical cross-entropy loss is defined in Eq.~\ref{eq:cross_entropy} where, $y_j$ denotes the ground-truth label for class $j$ and $\hat{y}_j$ is the predicted probability for class $j$ obtained from the softmax output of the model. The summation is taken over all $C$ classes.
\begin{equation}
\mathcal{L}_{\text{CE}}(Y, \hat{Y}) = -\sum_{j=1}^{C} y_j \log(\hat{y}_j)
\label{eq:cross_entropy}
\end{equation}
And, the KL divergence is given in Eq.~\ref{eq:kl_divergence}, where $p_j = \mathrm{softmax}(z_j^T / T)$ and $q_j = \mathrm{softmax}(z_j^S / T)$ represents the softened probability distributions from the teacher and student models, respectively.
\begin{equation}
\text{KL}(P \,\|\, Q) = \sum_{j=1}^{C} p_j \log\left(\frac{p_j}{q_j}\right)
\label{eq:kl_divergence}
\end{equation}

After optimizing using $\mathcal{L}_{\text{KD}}$, the refined global student model $\widetilde{\theta}_{M_S}^{(1)}$ is redistributed to all drones similar to Eq.~\ref{eq:init_model}, but for the current FL round $r_1$ (Eq.~\ref{eq:theta}): 
\begin{equation} \label{eq:theta}
    \theta_{M_S}^{i,(1)} \leftarrow \widetilde{\theta}_{M_S}^{(1)}    
\end{equation}

This completes round~$r_1$. For each subsequent round~$r_k$ ($k = 2, 3, \ldots, R$), the same process is repeated: each drone trains its local model using Eq.~\ref{eq:local_training}, sends updates to the server, which then aggregates them using Eq.~\ref{eq:fedavg}, performs KD as in Eq.~\ref{eq:kd_loss}, and redistributes the refined model to all drones. This iterative process continues for~$R$ total communication rounds, where the model is continually refined. The collaboration between FL and KD ensures that the final model generalizes well to unseen data and is robust against the heterogeneous, non-IID distributions across the drone swarm, while preserving privacy and minimizing communication overhead.

\subsection{Model Architecture} \label{modelarchitecture}
\begin{figure}[htbp]
\centering
\includegraphics[width=0.9\linewidth]{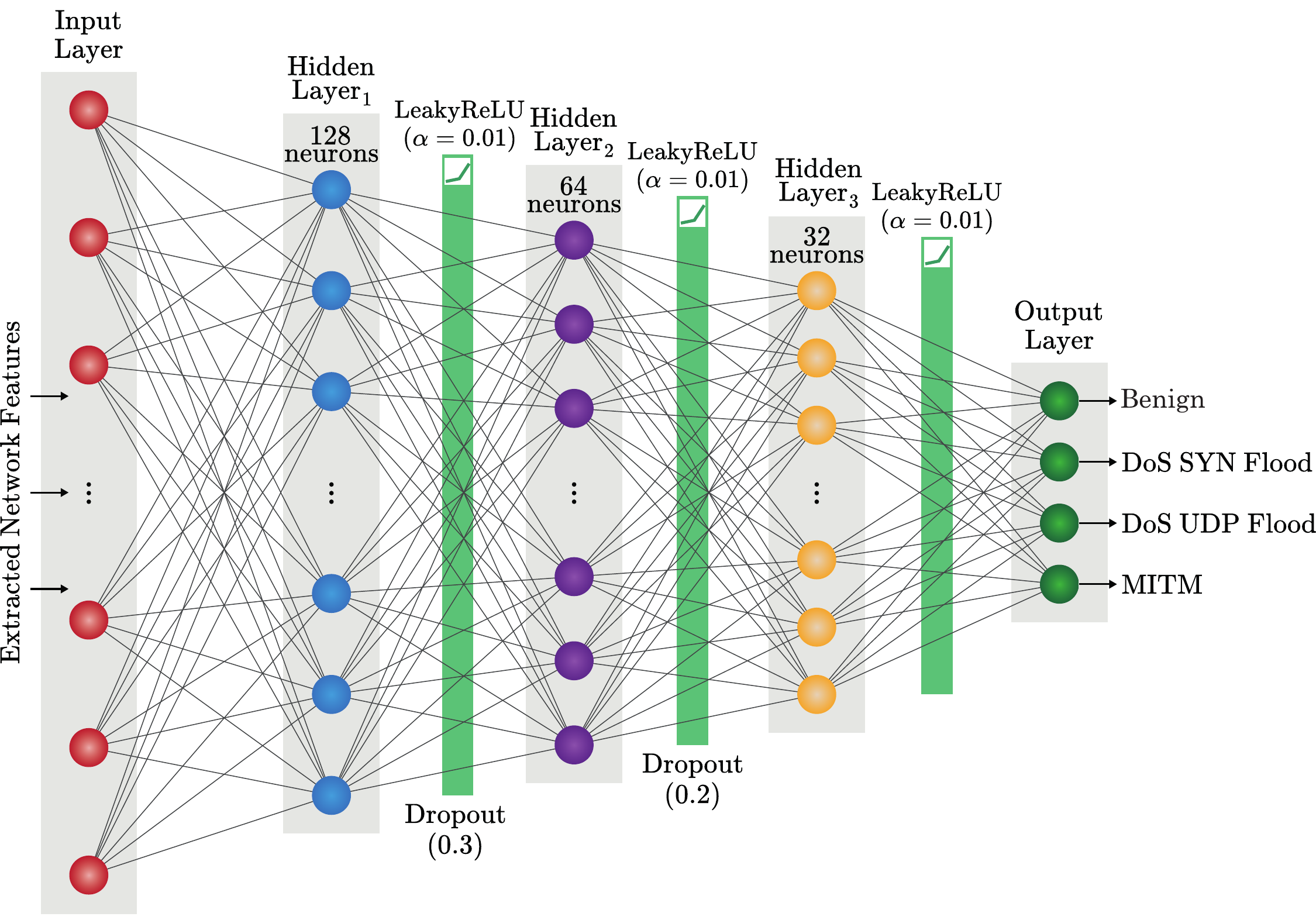}
\caption{Teacher model ($M_T$) architecture.}
\label{teacher_model}
\end{figure}

For drone swarm IDS, we prefer employing Deep Neural Network (DNN) -- one of the preferred methods for IDS due to its capability to process complex network traffic data and accurately identify abnormal behaviors~\cite{fu2025design, saad2024utilizing}. 
However, DNNs are computationally intensive and memory-demanding~\cite{li2024online}, and these requirements increase as model capacity and performance grow~\cite{sun2024logit}. Therefore, designing an efficient and powerful architecture is critical to ensure that the models can be deployed effectively on resource-constrained devices such as drones. 
To balance performance with resource efficiency, we propose a hierarchical architecture consisting of a high-capacity $M_T$ and lightweight student models, as illustrated in~Fig.~\ref{teacher_model}. The $M_T$ is used to guide the training of student models as the high-capacity reference model for server-side leveraging KD (KD is explained in details in Section~\ref{sec:KD}) so that smaller models to maintain high detection performance while being lightweight and suitable in an FL setup across resource-constrained drone networks. 

$M_T$ is pre-trained offline to achieve strong generalization and reliable guidance to the lightweight student models, consisting of a fully connected three hidden layers of 128, 64, and 32 neurons, respectively. Each hidden layer uses the LeakyReLU activation function to prevent neuron inactivation and improve gradient flow. Although the standard ReLU~($f(x) = \max(0, x)$) is widely used due to its simplicity and efficacy, it has a critical limitation known as dying ReLU, where neurons can become inactive and stop learning if it consistently output zero for negative inputs~\cite{bg2024comparative}. To mitigate this issue, we adopt Leaky ReLU, mathematically expressed as $f(x) = \max(\alpha x, x)$, where~$\alpha$ is a small positive constant (typically 0.01) used to allow a small non-zero gradient for negative inputs, thus maintaining some gradient flow even for inactive neurons, which improves network stability during training and higher final accuracy compared to ReLU~\cite{song2024comparative, bg2024comparative}. With these trade-offs, Leaky ReLU provides a suitable balance between computational efficiency, stability, and learning performance, making it an appropriate choice for IDS on resource-constrained drones.

The larger number of neurons allows the $M_T$ to capture complex feature patterns within the network traffic data, leading to richer representations. However, to further enhance generalization and prevent overfitting, dropping a fraction of neurons is an effective technique helps in reducing overfitting~\cite{yehezkel2024improving, pascual2024deep} that is strategically applied after the first and second hidden layers of the $M_T$ with dropout rates of 0.3 and 0.2, respectively. Wang et al.~\cite{wang2023data} show that moderate dropout on larger layers and skipping dropout on smaller ones results in strong regularization without losing essential features. The first hidden layer consists of 128 neurons, providing a higher capacity that increases the risk of overfitting; thus, a higher dropout rate is employed. The second hidden layer receives a slightly lower dropout rate as it has 64 neurons and to balance regularization and information retention. After the third hidden layer, which has 32 neurons, no dropout is applied to preserve critical feature representations essential for accurate classification. This selective strategy of dropout aligns with established practices, where larger layers benefit from higher dropout rates to mitigate overfitting, while smaller layers require less regularization to maintain performance. 

In the final classification layer, we employ a Softmax activation function (Eq.~\ref{eq:SoftMax}) for multi-class classification~\cite{mahima2023comparative} by producing a normalized probability distribution across the output classes~\cite{singh2023impact}. The Softmax activation function transforms input values into a probability distribution across four classes representing benign and different attacks by mapping them to values between 0 and 1 that sum to 1. Given an input vector~$\mathbf{z} = (z_1, z_2, \dots, z_K)$, the Softmax function is defined as:
\begin{equation}\label{eq:SoftMax}
\text{Softmax}(z_i) = \frac{e^{z_i}}{\sum_{j=1}^{K} e^{z_j}}
\end{equation}
where~$K$ is the number of classes,~$z_i$ is the input score for class~$i$, and the denominator ensures that all output probabilities sum to 1. Due to its exponential nature, the Softmax function emphasizes differences between input scores and assigns higher probabilities to larger inputs, which makes it effective for identifying the most probable class in multi-class problems.

The student model follows the overall structure of the $M_T$ but with reduced complexity to enable deployment on resource-constrained devices. Specifically, the three hidden layers are reduced to 64, 32, and 16 neurons, respectively, while retaining the same LeakyReLU activations and softmax output. This reduction significantly lowers the computational and memory demands while maintaining competitive performance, making the student model suitable for drone-based real-time intrusion detection.

\subsection{Model Optimization with Knowledge Distillation (KD)} \label{sec:KD}

Various model compression techniques have been explored by many researchers without losing much accuracy, including model quantization, model pruning, and KD methods~\cite{pang2024exploring}. 
Model pruning effectively eliminates redundant parameters and reduces model size by removing less important weights that do not impact neural network accuracy~\cite{kuzmin2023pruning} and quantization reduces model size and memory requirements by mapping high-precision floating-point weights (e.g., 32-bit) to lower-precision representations (e.g., 8-bit)~\cite{madnur2023enhancing}. 
Even though pruning and quantization are effective techniques for reducing model complexity, memory usage, and inference cost, they may also introduce slight performance degradation due to the potential removal or approximation of important weights~\cite{bao2024image}. 
KD is an effective compression method that can be used to transfer knowledge from a large and complex (teacher) model $M_T$ into a smaller and faster (student) model $M_S$ while maintaining a comparable performance~\cite{hinton2015distilling, bao2024image, dong2024ternary}. The main idea is to train the student model using both the hard labels (ground truth) and the soft labels produced by the teacher. After obtaining the logits from both models, a scaling factor called temperature $T > 1$ is applied to the softmax function to soften the predicted probability distribution across classes, producing soft labels that provide more information than hard labels~\cite{sun2024logit} using Eq.~\ref{eq:softmax_temp}:
\begin{equation}
p_i^{(T)} = \frac{\exp(z_i / T)}{\sum_j \exp(z_j / T)}
\label{eq:softmax_temp}
\end{equation}
where $z_i$ is the logit for class $i$, and $p_i^{(T)}$ is the softened output. These soft targets help to capture richer information about class relationships. For example, instead of predicting a single class with full confidence as in hard labels, soft labels can indicate that two classes share some similarity, rather than treating them as completely distinct, helping the student model to generalize better~\cite{sun2024logit}. The student is then trained to minimize a combined loss function consisting of two terms: a standard cross-entropy loss with the ground truth labels and a Kullback–Leibler (KL) divergence loss between the teacher and student softened outputs using Eq.~\ref{eq:kd_loss1}:
\begin{equation}
\mathcal{L}{\text{KD}} = (1 - \lambda) \cdot \mathcal{L}{\text{CE}}(\mathbf{y}, \hat{\mathbf{y}}S) + \lambda \cdot T^2 \cdot \mathcal{L}{\text{KL}}(\mathbf{p}_T^{(T)} \parallel \mathbf{p}_S^{(T)})
\label{eq:kd_loss1}
\end{equation}
where $\mathbf{y}$ is the ground truth label, $\hat{\mathbf{y}}_S$ is the student's predicted distribution, $\mathbf{p}_T^{(T)}$ and $\mathbf{p}_S^{(T)}$ are the teacher and student outputs softened with temperature $T$, and $\lambda \in [0, 1]$ balances the contribution of the two losses. The factor $T^2$ is used to properly scale the gradients coming from the distillation loss.

\section{Experimental Analysis} \label{sec:results}

\subsection{Dataset}
To evaluate the proposed FL-KD framework for drone swarm network IDS, we utilized the large-scale ISOT Drone Dataset~\cite{chen2025drone}, which is a real-world dataset collected from DJI Tello Edu drones that contains both benign and attack scenarios to reflect realistic drone network traffic over standard IP-based communication protocols, including TCP and UDP flows. This is consistent with drone communication architectures that leverage wireless IP-based connectivity (e.g., WiFi or LTE)~\cite{chavekar2023review, mushtaq2024framework}. The dataset includes more than 2.8 million samples extracted from 14 hours of anomalous and 10 hours of benign network activity, which were captured in PCAP format and then extracted into CSV files. We label the data using its distinct folder structure (e.g., Benign, MITM, DoS). For our experiments, we focused on anomalous traffic types highly relevant to drone security, such as TCP SYN Flood (as DDoS), UDP Flood, and MITM attacks. 

For an effective evaluation, data selection process was as follows. The attack scenarios with very limited data were excluded, as maintaining reasonable class representation is critical in FL to ensure robust and unbiased model convergence. The selected malicious scenarios simulate attacks targeting drone communication vulnerabilities such as (a) TCP SYN Flood attacks (about 3.5 hours long attacks with 295,436 samples), aiming to overwhelm critical drone communication services with massive SYN requests, exhausting network resources, and causing DoS within drone command and control systems; (b) UDP flooding attacks (approximately 5 hours long, with 881,497 samples) flooding drones with a large volume of UDP traffic, leading to communication congestion or complete disruption; and (c) MITM attacks (roughly 1.5 hours long with 96,918 samples) intercepting communications between drones and GCS, compromising the confidentiality and integrity of mission-critical data. 

\subsubsection{Data Preprocessing}

As an initial step in data preprocessing, we cleaned irrelevant features such as timestamps from the dataset (ISOT Drone dataset~\cite{chen2025drone}). Temporal features were intentionally deselected to align with the objective of IDS to focus on communication patterns and attack characteristics. 
Next, all numerical features were normalized using min-max scaling, mapping values into the [0, 1] range, to prevent features with larger numeric ranges from dominating the learning process and also to preserve the original distribution shape of each feature for a faster and more stable convergence. 

\begin{figure}[htbp]
\centering
\begin{subfigure}[t]{0.48\textwidth}
    \centering
    \includegraphics[width=\linewidth]{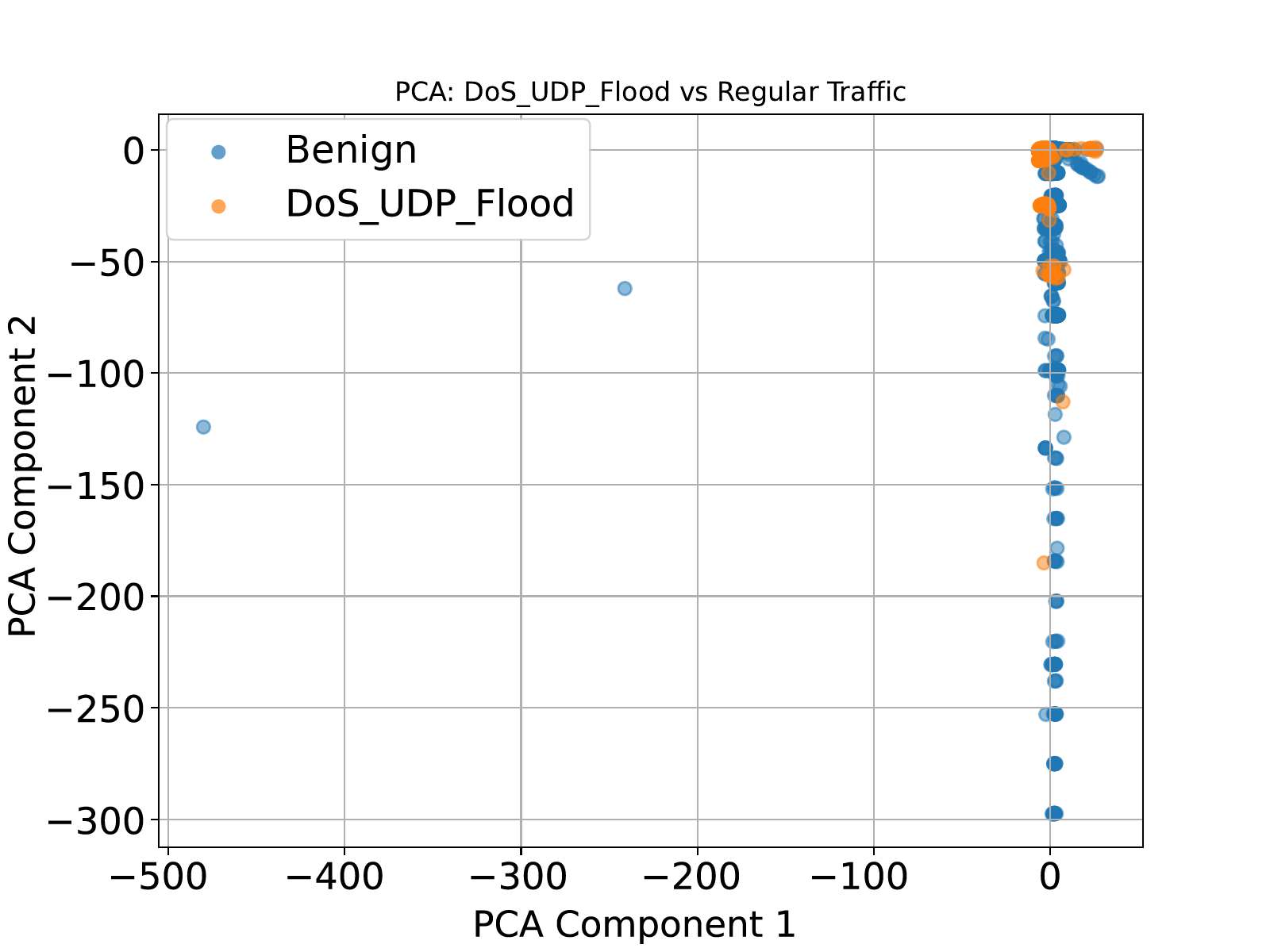}
    \caption{UDP flooding attack.}
    \label{pca_udp}
\end{subfigure}
~
\begin{subfigure}[t]{0.48\textwidth}
    \centering
    \includegraphics[width=\linewidth]{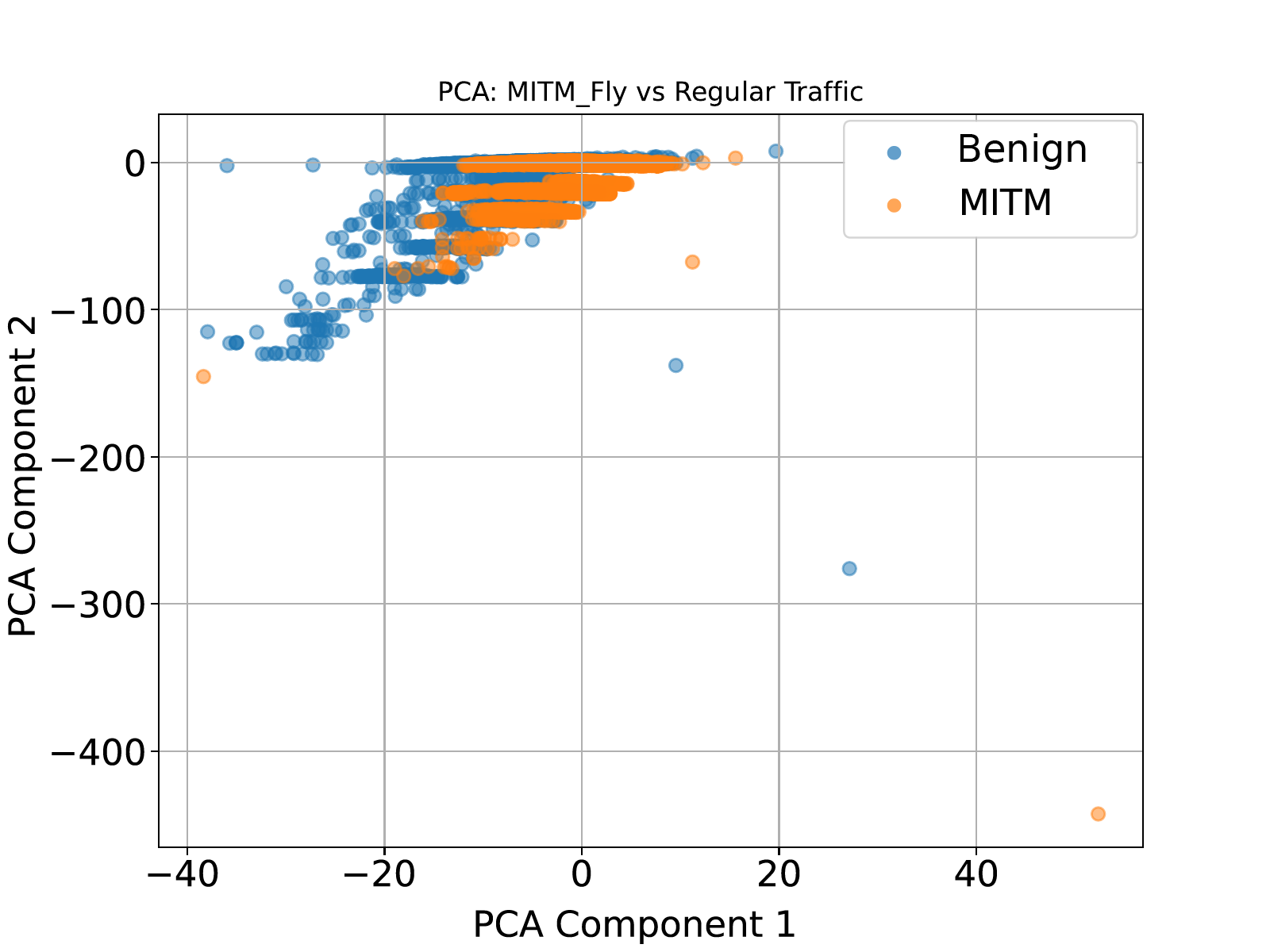}
    \caption{MITM attack.}
    \label{pca_mitm}
\end{subfigure}
\caption{PCA for (a) UDP flooding attack (DoS attack) and (b) MITM attack vs. benign data.}
\label{pca_udp_mitm}
\end{figure}

We also applied PCA as an exploratory step to reduce the dataset into lower dimensions and to identify hidden patterns and behavior of different traffic classes by capturing maximum variances in the data~\cite{zyad2024evaluation}. 
The PCA results showed that some attack classes deviated significantly from benign traffic, while others exhibited notable overlap, highlighting the complexity of the classification task. For instance, Fig.~\ref{pca_udp} illustrates the distribution of the DoS UDP Flood attack against benign traffic with substantial overlap, suggesting challenges in clearly separating these two classes. In contrast, the MITM attack illustrated in Fig.~\ref{pca_mitm} exhibited a more noticeable deviation from the benign traffic cluster, implying higher separability. Based on these findings, we retained the meaningful features after preprocessing to preserve critical patterns necessary for effective classification. Overall, the PCA analysis validates our selection of attack scenarios by demonstrating a range of deviations, from easily detectable to difficult to detect behaviors, thereby providing a comprehensive evaluation of intrusion detection capabilities in drone networks.

To further optimize the feature set and improve model efficiency, two correlation analyses were conducted. First, feature-to-feature correlation was evaluated using the Pearson Correlation Coefficient (PCC)~\cite{putro2024feature} across the entire dataset to identify highly correlated features. PCC measures the relationship between two features, as defined in Eq.~\ref{eq:pcc}, 
\begin{equation}
r = \frac{\sum (x_i - \bar{x})(y_i - \bar{y})}{\sqrt{\sum (x_i - \bar{x})^2 \sum (y_i - \bar{y})^2}}
\label{eq:pcc}
\end{equation}
where $x_i$ and $y_i$ are individual values of the two features being compared, and $\bar{x}$ and $\bar{y}$ are their respective means. A correlation value close to $\pm1$ indicates a strong relationship, while values near $0$ suggest weak to no correlation. In our analysis, feature pairs with an absolute PCC greater than 0.85 were considered highly correlated, and one feature from each pair was removed to reduce redundancy and improve model efficiency. 
However, to decide which feature to retain among correlated pairs, a feature-to-class correlation analysis was performed using the Point-Biserial Correlation Coefficient (PBCC), which is a method designed to measure the strength of association between a continuous feature and a binary target variable~\cite{sibarani2024relationships}. PBCC evaluates how differently a feature behaves across the two classes by comparing the means of that feature for the benign and attack samples. The PBCC can be defined by the Eq.~\ref{eq:pbcc}:
\begin{equation}
r_{pb} = \frac{M_1 - M_0}{s} \sqrt{ \frac{n_1 n_0}{n^2} }
\label{eq:pbcc}
\end{equation}
where $M_1$ and $M_0$ are the means of the continuous features for the attack and benign classes, $s$ is the standard deviation of the feature across all samples, and $n_1$, $n_0$, and $n$ are the number of samples in each class and the total number of samples, respectively. This analysis was used to quantify how relevant each feature is for distinguishing between benign and attack traffic. Features demonstrating stronger relevance to attack detection were retained, while features with weak association were removed. These combined correlation-based steps ensured that the final feature set was both compact and highly informative, enabling efficient and robust learning in federated drone network environments.

We partitioned our dataset into two subsets: 20\% for the construction of the proxy dataset for training the teacher model with no client overlap, and the remaining 80\% for the distributed (five) clients, each representing a distinct drone behavior. Each class was assigned to clients with percentage variations, approximately between 23\% and 27\%, to ensure a realistic non-IID distribution. Each client's assigned data was then further divided into 80\% for training and 20\% for validation to reflect real-world heterogeneity.

For the experimentation of proposed FL-based IDS framework, we designed a physical testbed consisting of five Raspberry Pi 4 Model B (quad-core ARM Cortex-A72 CPU operating at 1.5 GHz and 4 GB of LPDDR4 RAM, running the Raspbian GNU/Linux 11 (Bullseye)) as resource-constrained drone clients and a MacBook Pro (3.0 GHz dual-core Intel Core i7 processor with 16 GB of RAM) serving as the central federated server. 
This hardware configuration provides a realistic environment reflecting practical limitations on computational resources, memory capacity, and communication bandwidth typically found in drone swarm networks as Raspberry Pi is a typically chosen mission computer for drones~\cite{tufekci2024dude}. 
For software specifications, we utilized Python and TensorFlow for model development and training, and the Flower framework to manage the FL process across the distributed clients and server.

\subsubsection{Experimental Results}
The experimental evaluation covers the security performance and the efficiency of system and communication resources in terms of resource usage and network overhead. All experiments were conducted over ten FL rounds to ensure a consistent and fair comparison. The FL-Baseline refers to the traditional FL framework without incorporating KD, using the larger DNN model (i.e., the teacher model), while the FL-KD setup integrates the KD technique into the federated process using the smaller DNN model (i.e., the student model) as described in Section~\ref{modelarchitecture}.

\begin{figure}[htbp]
\centering
\begin{subfigure}[t]{0.48\textwidth}
    \centering
    \includegraphics[width=.95\columnwidth]{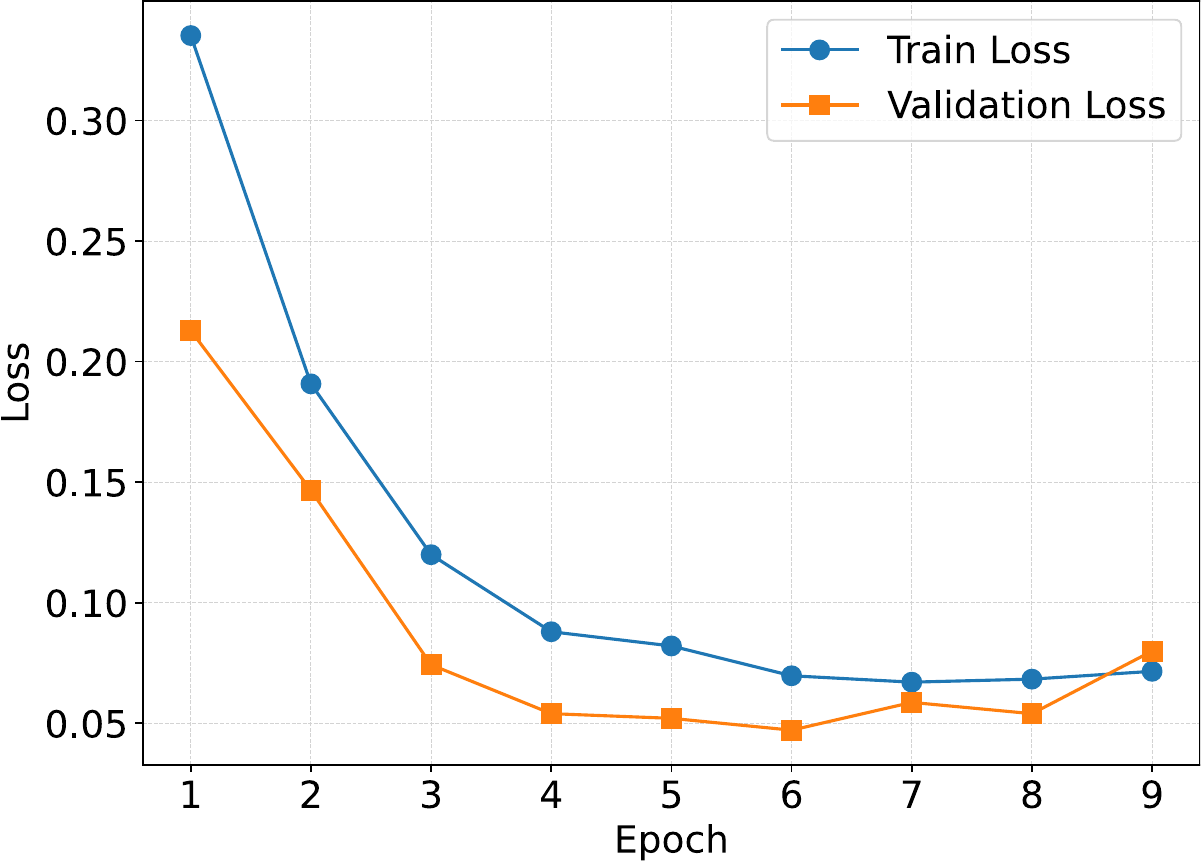}
    \caption{Pre-trained teacher model: training and validation loss.}
    \label{teacher_loss}
\end{subfigure}
~
\begin{subfigure}[t]{0.48\textwidth}
    \centering
    \includegraphics[width=\columnwidth]{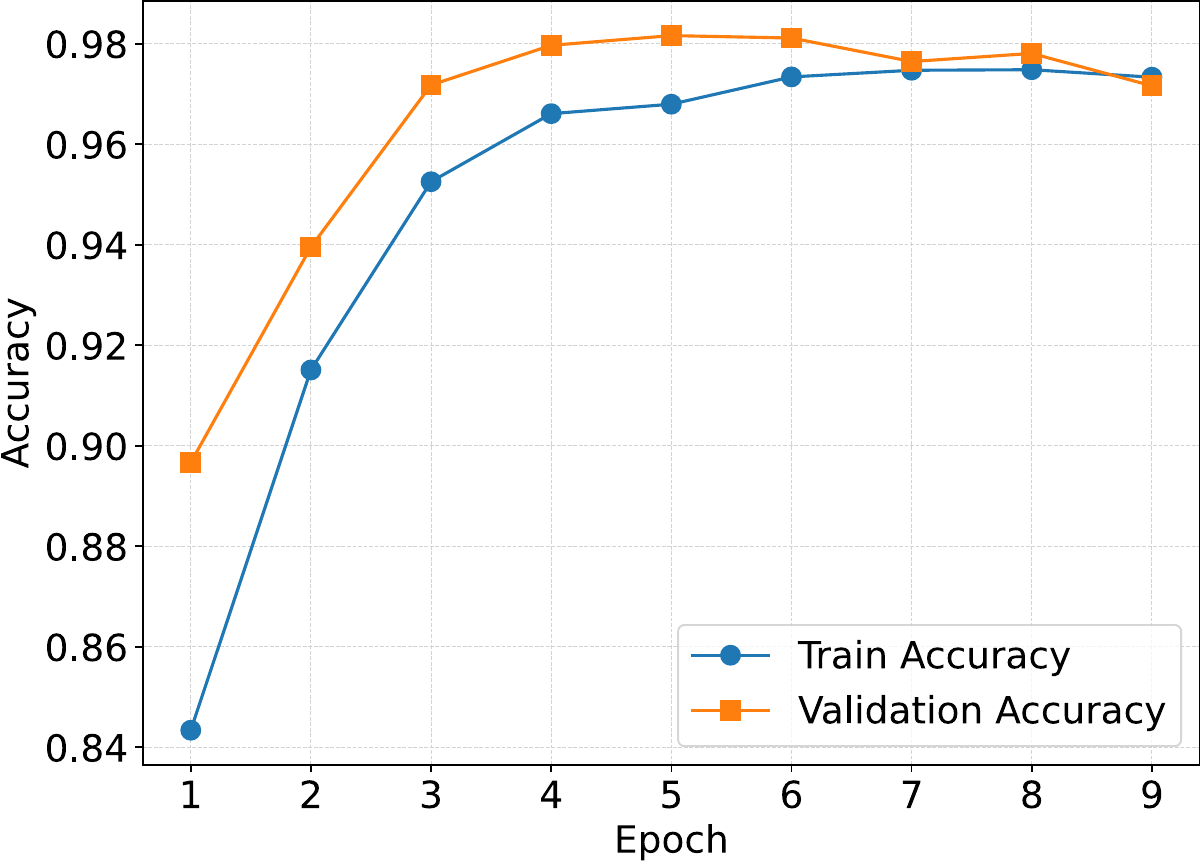}
    \caption{Pre-trained teacher model: training and validation accuracy.}
    \label{teacher_acc}
\end{subfigure}
\caption{Pre-trained teacher model: (a) training and validation loss, (b) training and validation accuracy.}
\label{teacher_loss_acc}
\end{figure}

First, we evaluate the performance of the pre-trained teacher model used in the KD process. The pre-trained teacher model achieved a validation accuracy of approximately 98\% and a validation loss below 0.05, demonstrating strong generalization performance. As shown in Fig.~\ref{teacher_loss}, the training and validation loss initially start around 0.3 and 0.2, respectively, and decrease gradually as training progresses, eventually stabilizing close to 0.05 during the later epochs, indicating effective convergence.
Similarly, both the training and validation accuracy, illustrated in Fig.~\ref{teacher_acc}, gradually improve during the training process, reaching close to 98\% with minimal gap between them, confirming the robustness and generalization capability of the teacher model. These results demonstrate the suitability of the teacher model for guiding the student models via KD.

\begin{figure}[htbp]
\centering
\begin{subfigure}[t]{0.48\textwidth} 
    \centering
    \includegraphics[width=\columnwidth]{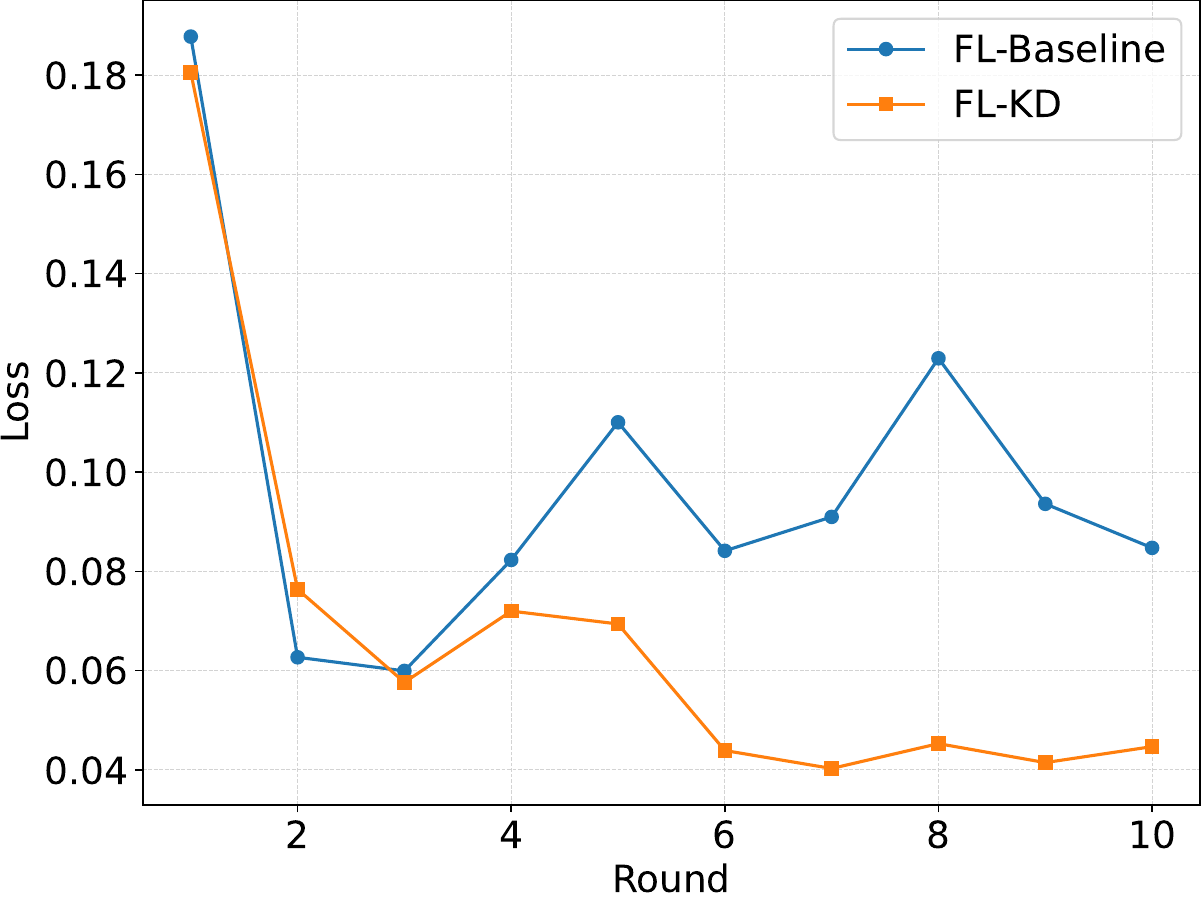}
    \caption{Global student model loss.}
    \label{gsm_loss}
\end{subfigure}
~
\begin{subfigure}[t]{0.48\textwidth} 
    \centering
    \includegraphics[width=\columnwidth]{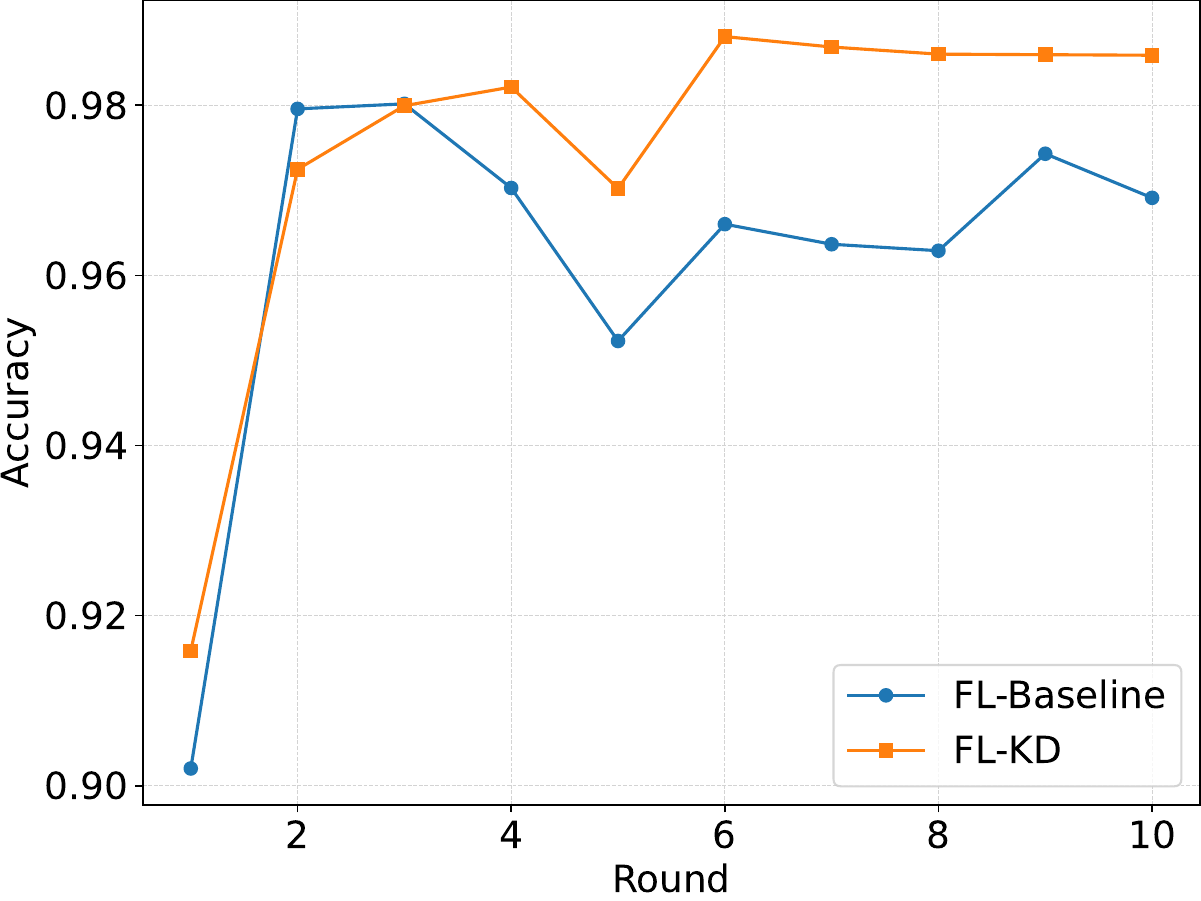}
    \caption{Global student model accuracy.}
    \label{gsm_accuracy}
\end{subfigure}
\caption{Global student (a) model loss and (b) accuracy for FL-Baseline vs. FL-KD comparison.}
\label{gsm_loss_acc}
\end{figure}

Fig.~\ref{gsm_loss} and Fig.~\ref{gsm_accuracy} illustrate the comparison of training loss and accuracy across FL rounds for the global student model trained under the FL-Baseline and FL-KD frameworks.
In the FL-Baseline framework, the global student model achieved a rapid improvement during the early rounds, with the validation accuracy rising from approximately 90\% in round 1 to about 97\% by round 10. The corresponding validation loss decreased gradually from around 0.18 to approximately 0.08. In contrast, the FL-KD framework demonstrated stronger performance, with the validation accuracy improving from approximately 92\% in round 1 to about 98.6\% by round 6, and then stabilizing at a high level through round 10. The validation loss in FL-KD also decreased and reached around 0.04 during the later rounds. These results clearly demonstrate that KD improves both the convergence speed and the final performance of the global student model, resulting in higher accuracy and lower loss compared to the baseline without KD.

\begin{figure}[htbp]
\centering
\begin{subfigure}[t]{0.48\textwidth} 
    \centering
    \includegraphics[width=\columnwidth]{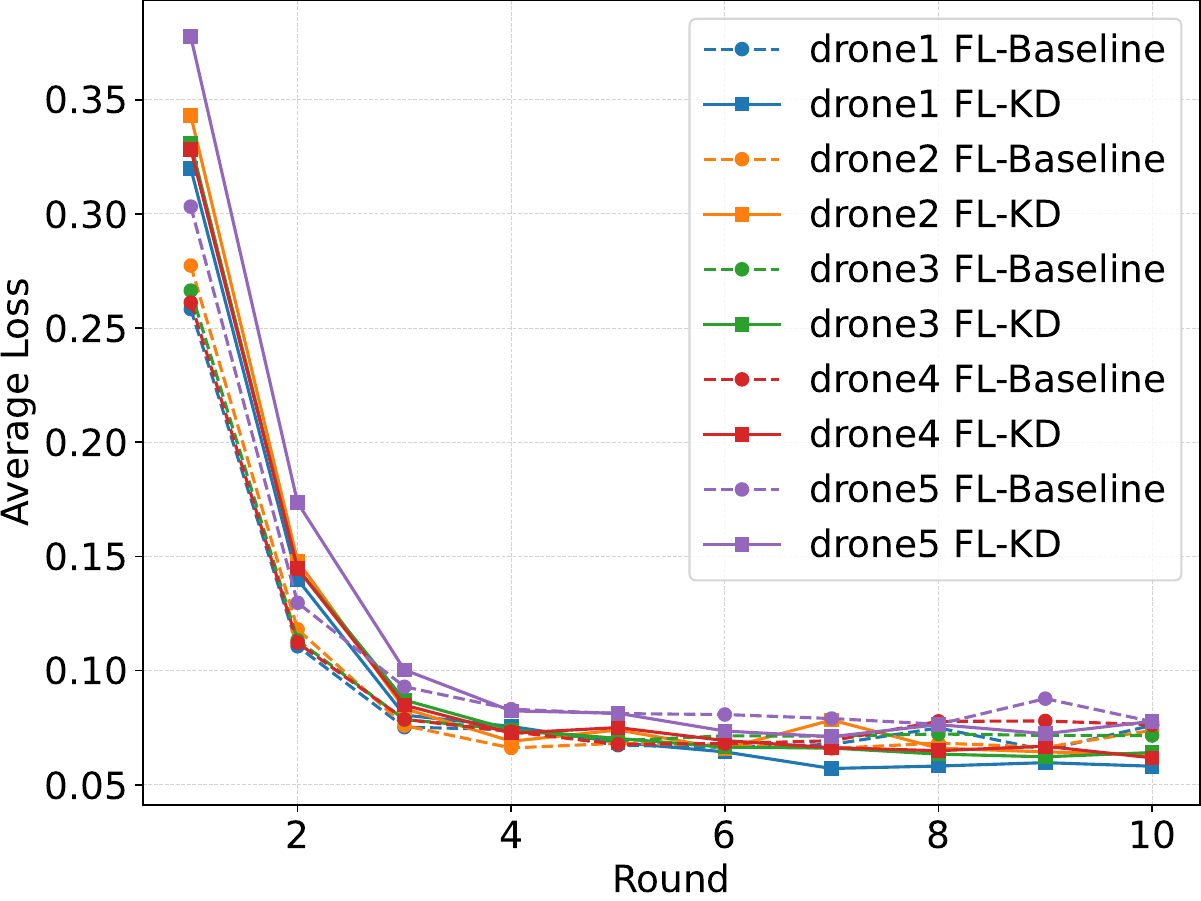}
    \caption{Training loss across drones: FL-Baseline vs. FL-KD.}
    \label{local_loss}
\end{subfigure}
~
\begin{subfigure}[t]{0.48\textwidth} 
    \centering
    \includegraphics[width=\columnwidth]{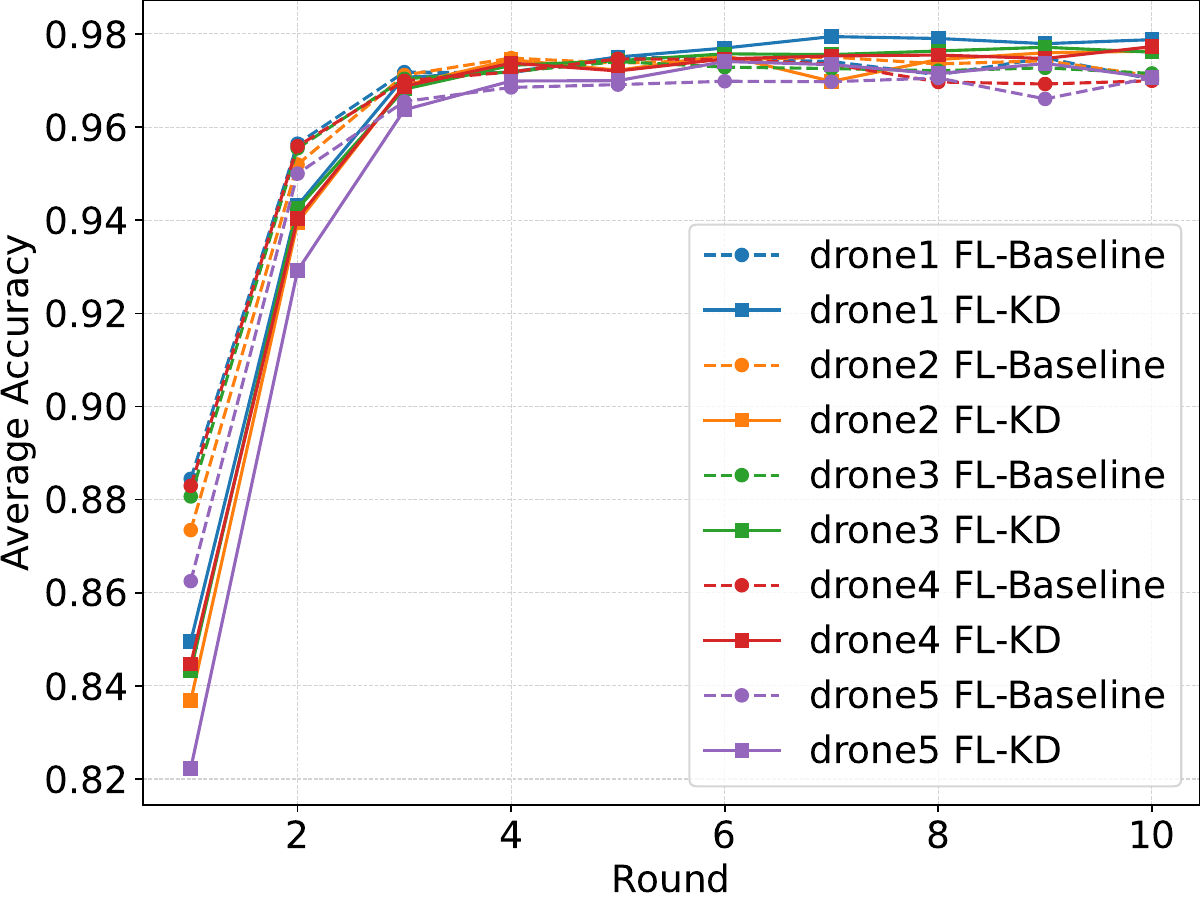}
    \caption{Training accuracy across drones: FL-Baseline vs. FL-KD.}
    \label{local_accuracy}
\end{subfigure}
\caption{(a) Training loss and (b) accuracy across drones comparing FL-Baseline and FL-KD.}
\label{local_loss_acc}
\end{figure}

The performance of the local student models on each drone client was evaluated separately for both the FL-Baseline and FL-KD. Fig.~\ref{local_loss} and Fig.~\ref{local_accuracy} illustrate the training loss and accuracy per round for all drones. It is worth noting that the student models used in FL-KD are smaller than those used in FL-Baseline, yet they achieve faster convergence and higher final accuracies. In the FL-Baseline, the initial training accuracy across drones ranged between approximately 86\% and 88\% during round 1, quickly improved to 93\%-95\% by the second round, and gradually improved to values between 96\% and 97\% by round 10. The corresponding training loss decreased steadily from values around 0.26~-~0.30 in round 1 to values between 0.07 and 0.08 at round 10. In contrast, the FL-KD demonstrated faster convergence and higher final accuracies. The initial training accuracies for FL-KD started slightly lower for some drones, ranging from approximately 82\% to 85\% in round 1, but improved more rapidly, reaching accuracies of around 97\% to 98\% by round 10. Similarly, the training loss under FL-KD decreased gradually and reached close to 0.06 or below for most drones by the end of training. These results indicate that KD not only improves the final detection performance at the local clients but also accelerates the convergence of the student models during training.

We also compared the model size used by the clients in both FL-Baseline and FL-KD to evaluate the resource efficiency improvement achieved by the KD approach. As shown in Table~\ref{tab:model_size}, the student models used in the FL-KD have a significantly smaller size compared to those used in the FL-Baseline, decreasing from approximately 0.23 MB to 0.11 MB, representing a reduction of about 52.6\%. The substantial reduction, which aligns with the architectural design of the student models, demonstrates the effectiveness of the KD technique in enabling lightweight models suitable for resource-constrained drone environments, while still maintaining strong IDS performance.

\begin{table}[!t]
\caption{Model Size Comparison between FL-Baseline and FL-KD}
\label{tab:model_size}
\small
\centering
\begin{tabularx}{\columnwidth}{l X c}
\hline
\textbf{Model} & \textbf{Type} & \textbf{Size (MB)} \\
\hline
FL-Baseline & Larger (Teacher) & 0.23 \\
FL-KD       & Smaller (Student) & 0.11 \\
\hline
\end{tabularx}
\end{table}

\begin{figure}[htbp]
\centering
\begin{subfigure}[t]{0.48\textwidth} 
    \centering
    \includegraphics[width=\linewidth]{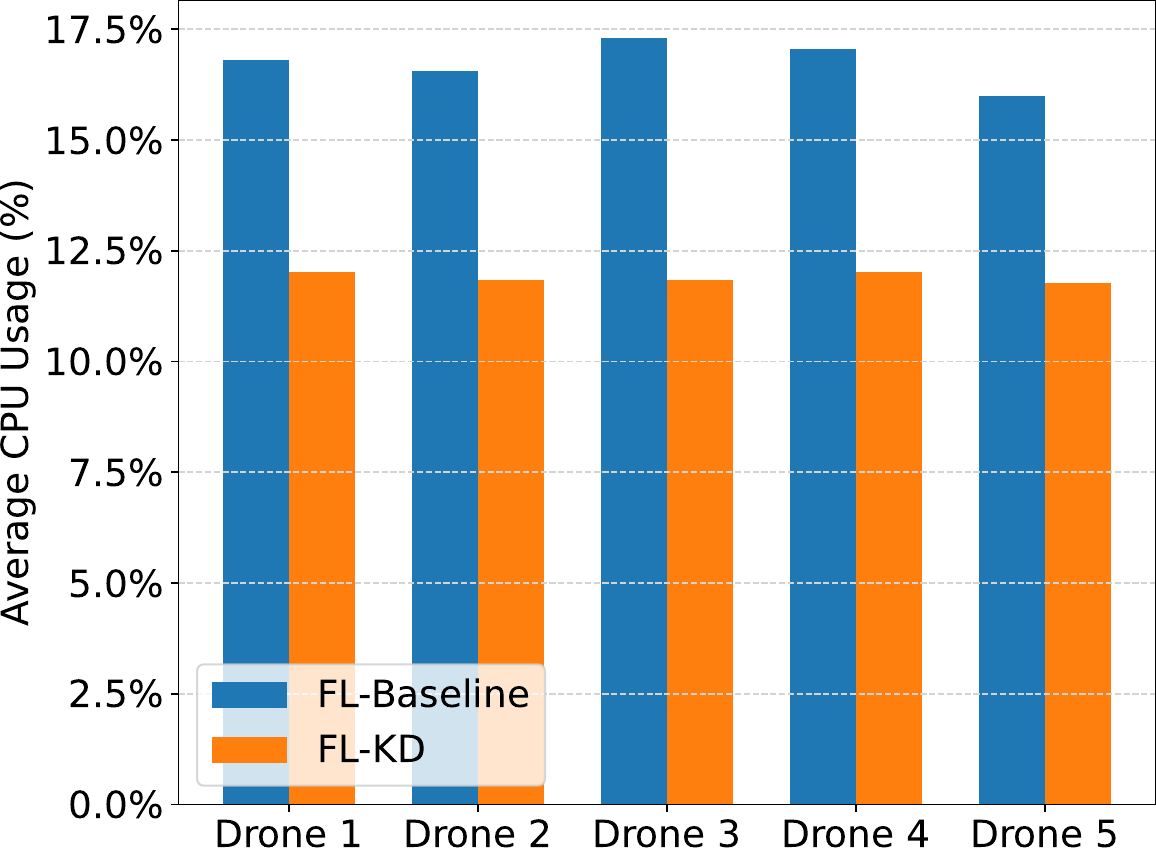}
    \caption{Average CPU usage comparison.}
    \label{cpu_usage}
\end{subfigure}
~
\begin{subfigure}[t]{0.48\textwidth} 
    \centering
    \includegraphics[width=\linewidth]{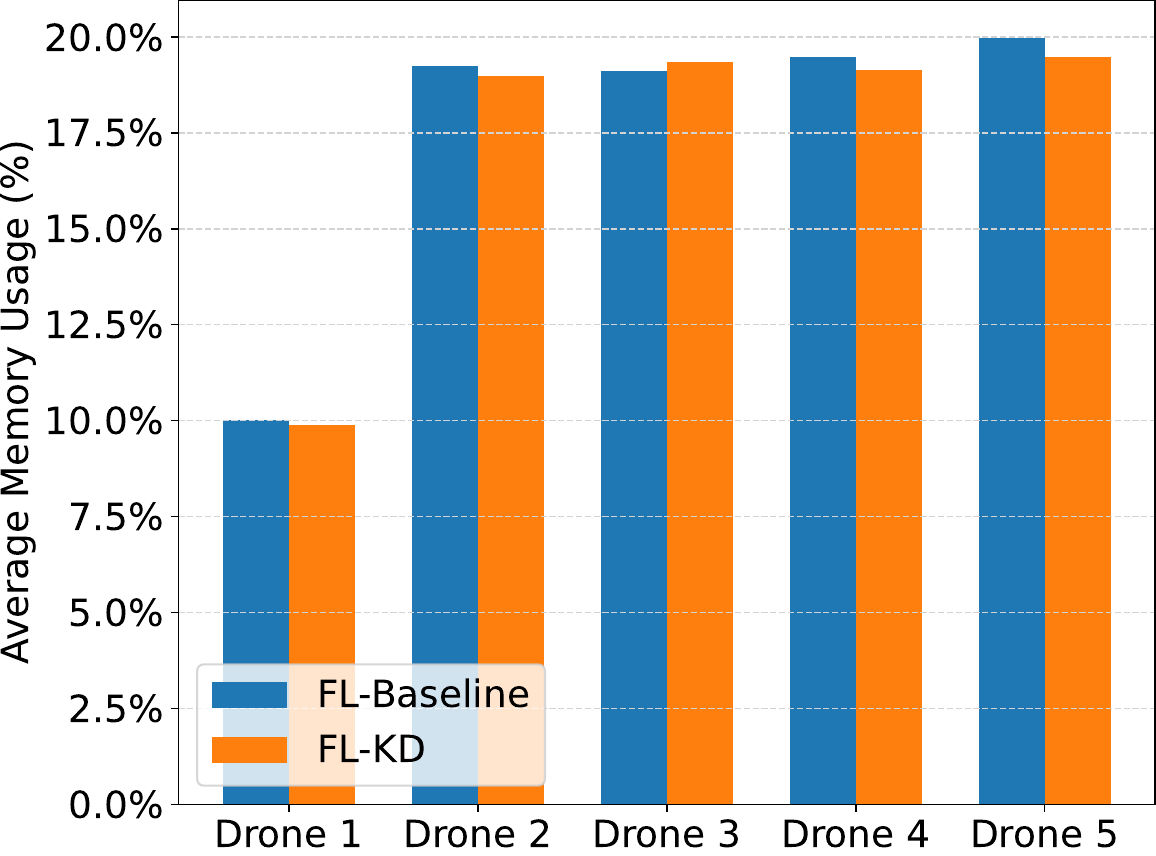}
    \caption{Memory usage comparison.}
    \label{memory_usage}
\end{subfigure}
\caption{FL-Baseline vs. FL-KD CPU and memory usage comparison across drones.}
\label{cpu_memory_usage}
\end{figure}

As drones have limited computational capabilities, we also compared the CPU and memory utilization of the FL-Baseline and FL-KD methods. 
For the CPU utilization, we monitored each drone throughout the training process.  
As shown in~Fig.~\ref{cpu_usage}, the average CPU usage for all drones under the FL-KD framework is consistently lower compared to the FL-Baseline framework. Specifically, the average CPU usage across all drones was reduced from 17\% in FL-Baseline to 12\% in FL-KD, representing an approximate 29\% reduction. This reduction can be attributed to the smaller and lightweight student models used in FL-KD, which require fewer computational resources during local training. By contrast, the larger models employed in the FL-Baseline framework result in higher CPU utilization. These results confirm that the KD approach not only improves model efficiency but also significantly reduces the computational cost on resource-constrained drone devices. 
In addition to the CPU utilization, the FL-KD resulted in slightly lower average memory consumption across most drones compared to FL-Baseline as illustrated in~Fig.~\ref{memory_usage}. For example, Drone 2 showed a reduction from 19.32\% to 18.97\%, while Drone 4 decreased from 19.57\% to 19.13\%. However, the overall difference was marginal, with all variations falling below half a percent. While CPU usage decreased significantly with FL-KD due to the reduced number of computations, memory usage demonstrated only minor improvements because the dominant memory consumption arises not only from model parameters, but also from static allocations by the runtime environment, cached data, and pre-allocated training buffers. These components tend to remain stable regardless of model size, which limits the impact of using smaller student models on overall memory usage.

\begin{figure}[htbp]
\centering
\includegraphics[width=0.6\columnwidth]{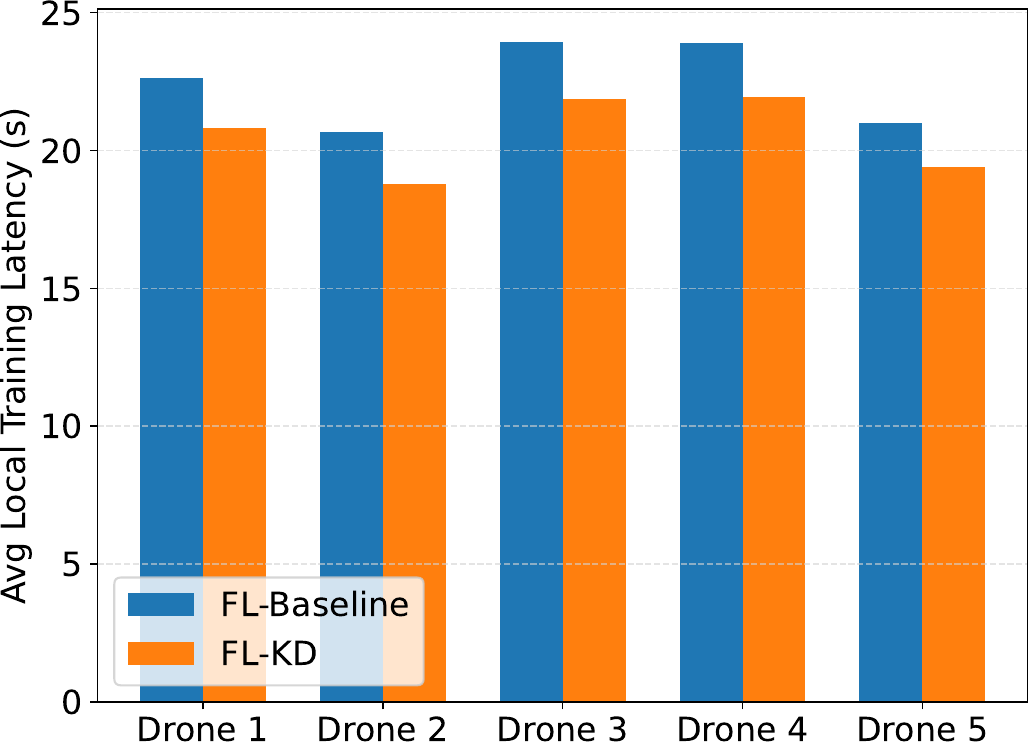}
\caption{Local training latency: FL-Baseline vs. FL-KD.}
\label{local_latency}
\end{figure}

In addition to model accuracy and resource efficiency, we evaluated the local training latency per client to measure the computational cost experienced by individual drones. As shown in~Fig.~\ref{local_latency}, the proposed FL-KD framework consistently achieved lower model training latency across all drones compared to the FL-Baseline. Specifically, average latency reductions ranged between 1.5s and 2.1s per round. For example, Drone 1 experienced a decrease from an average of 21.10s in the FL-Baseline to 19.00s under FL-KD, while Drone 2 improved from 19.00s to 17.00s. These reductions are due to the smaller and more efficient student models used in FL-KD, which require fewer computations per training round. Across all five drones, the mean latency decreased from 20.64s in FL-Baseline to 18.90s in FL-KD with a total improvement of approximately 9.4\%. This finding further demonstrates the practicality of KD in federated drone networks, where reduced training latency enhances responsiveness and preserves battery life without compromising detection performance.

\begin{figure}[hbt]
\centering
\begin{subfigure}[t]{0.48\textwidth} 
    \includegraphics[width=\linewidth]
    {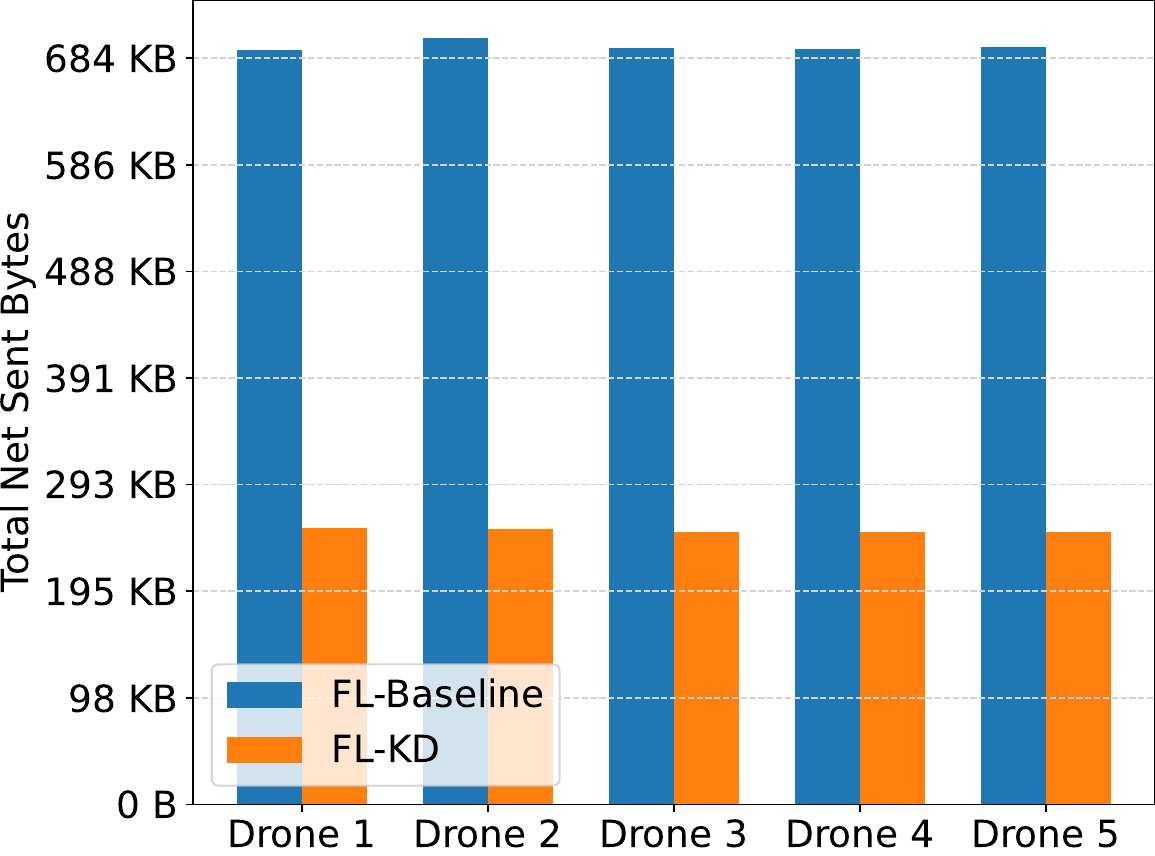}
    \caption{Sent bytes.}
    \label{fig:net_sent_bytes}
    \end{subfigure}
~
\begin{subfigure}[t]{0.48\textwidth} 
    \includegraphics[width=\linewidth]{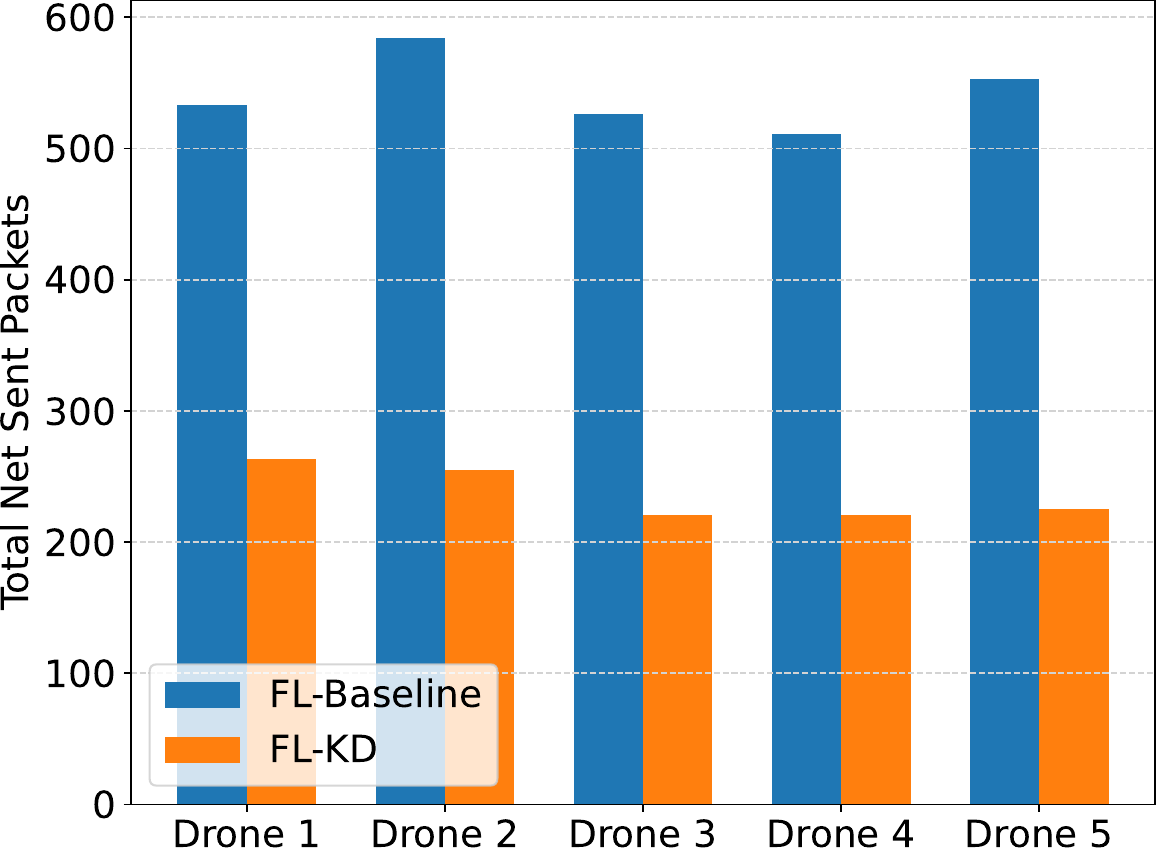}
    \caption{Sent packets.}
    \label{fig:net_sent_packets}
\end{subfigure}
\hfill
\begin{subfigure}[t]{0.48\textwidth}  
    \includegraphics[width=\linewidth]{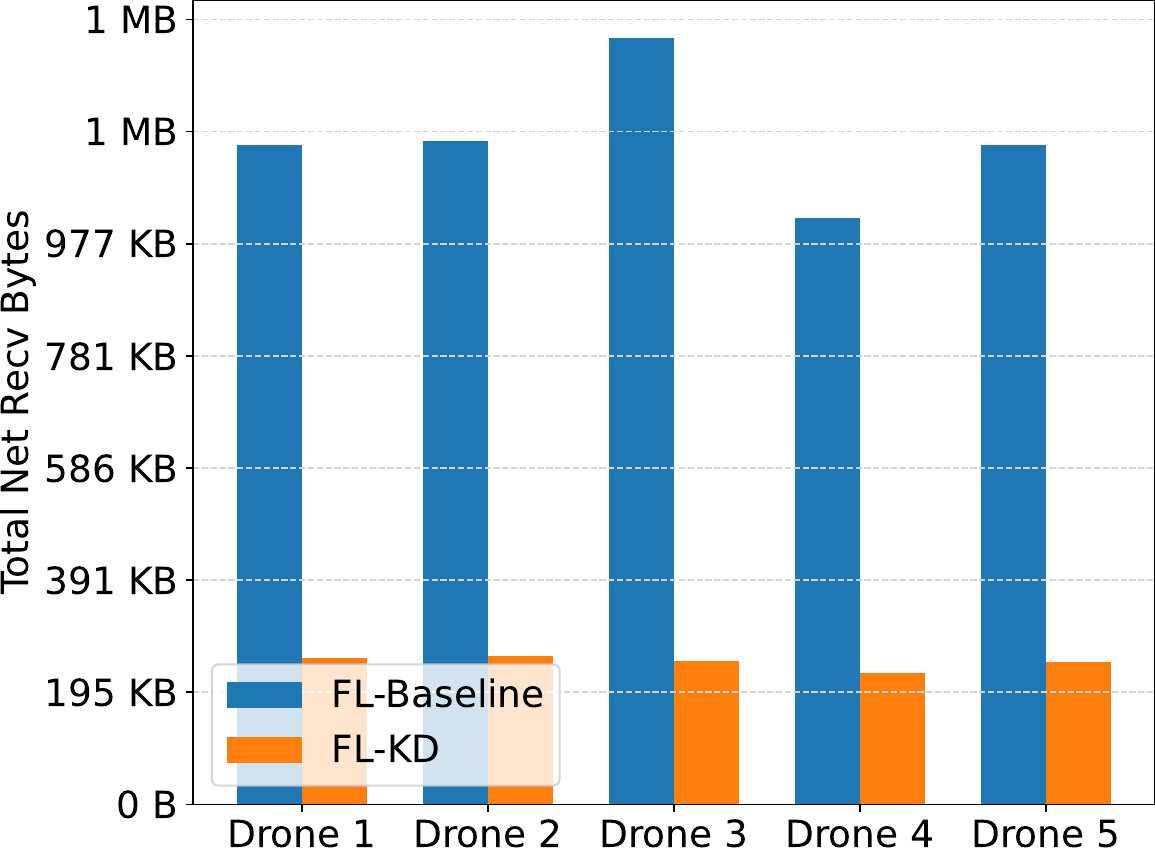}
    \caption{Received bytes.}
    \label{fig:net_recv_bytes}
\end{subfigure}
~
\begin{subfigure}[t]{0.48\textwidth}  
    \includegraphics[width=\linewidth]{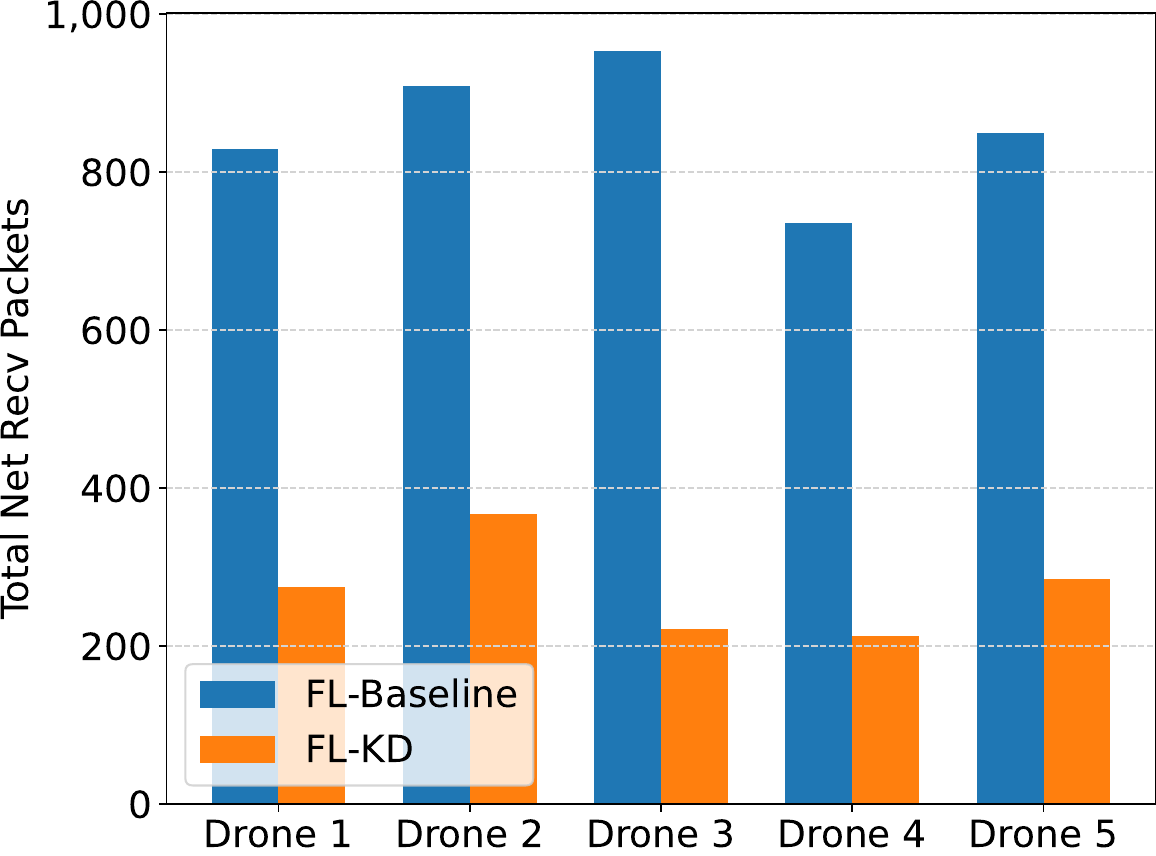}
    \caption{Received packets.}
    \label{fig:net_recv_packets}
\end{subfigure}
\caption{Network communication comparison between FL-Baseline and FL-KD.}
\label{fig:network_all}
\end{figure}

Another important indicator for the applicability of the proposed approach is the bandwidth usage. As drones are mobile and depend on wireless communication, high bandwidth or reliable communication may not be always available. Therefore, for a more reliable solution, the network usage (both sent/received packets/data size) should be minimized. 
Our observations show that the FL-KD framework achieved a substantial reduction in network communication compared to FL-Baseline. For the total sent bytes illustrated in~Fig.~\ref{fig:net_sent_bytes}, each drone in FL-Baseline transmitted approximately 690–700 KB across 10 rounds. This was reduced by approximately 64\% to 250 KB per drone in FL-KD. Similarly, the number of sent packets in FL-KD was almost half or less compared to FL-Baseline, representing an average reduction of around 56\%, as shown in~Fig.~\ref{fig:net_sent_packets}, indicating fewer transmissions and a lighter communication load. 
A similar trend was observed for received bytes as illustrated in~Fig.~\ref{fig:net_recv_bytes}, where each drone received around 1 MB of model updates in FL-Baseline, while the total received data per drone dropped by approximately 76\% to 250 KB in FL-KD. The number of received packets also decreased substantially with an average reduction of around 68\% as shown in~Fig.~\ref{fig:net_recv_packets}, confirming the significant reduction in overall communication overhead.

These findings underscore more than just a reduction in communication overhead and highlight a critical interaction in drone swarm systems, where communication and computation are closely intertwined affecting each other's performance~\cite{wang2019survey}. Communication enables the exchange of model updates and detection outcomes, while computation processes this data for real-time decision-making. In drone swarm environments requiring real-time responsiveness, such as intrusion detection and dynamic threat mitigation or even for tasks including but not limited to object detection, tracking, collusion avoidance, high-latency communication can delay model updates, reduce detection accuracy, and create bottlenecks that limit timely decision-making. Therefore, enhancing communication efficiency and reducing decision latency have a direct positive impact on the computational performance of the swarm and accelerating model updates, and improving detection reliability. Our FL-KD framework addresses this by reducing both the size and frequency of data transmissions through lightweight student models and server-side distillation. 
Our work achieves higher detection accuracy, faster convergence, reduced computational load, slightly improved memory usage, and substantially lower network communication overhead compared to the FL-Baseline, demonstrating the suitability of FL-KD for deployment in resource-constrained drone swarm environments, where both security and communication efficiency are critical requirements.

\section{Conclusion} \label{sec:conclusion}
While drones and drone swarms are gaining significant interest across various domains, their security remains underdeveloped. Traditional centralized intrusion detection methods are not well-suited for drone swarm network security due to the high communication and computational overhead they impose. To address these challenges, this paper introduces a lightweight federated learning (FL)-based intrusion detection system (IDS) tailored specifically for drone swarm networks. By incorporating knowledge distillation (KD), the proposed framework enables individual drones to train and utilize compact models that maintain high detection accuracy while minimizing resource consumption. We evaluated our approach using Raspberry Pi 4 devices (for drone mission computers) and a real-world drone network dataset. Our experimental results demonstrate the effectiveness of our proposed approach, achieving up to 98.6\% detection accuracy, a 70\% reduction in communication overhead, and a 29\% decrease in computational load compared to conventional FL-based methods. These results demonstrate the feasibility and practicality of deploying a secure, efficient, and scalable FL-KD IDS tailored for real-time drone networks.

\bibliographystyle{elsarticle-num} 
\bibliography{references}

\end{document}